\documentclass[sigplan,screen]{acmart}
\AtBeginDocument{%
  \providecommand\BibTeX{{%
    \normalfont B\kern-0.5em{\scshape i\kern-0.25em b}\kern-0.8em\TeX}}}

\setcopyright{acmcopyright}
\copyrightyear{2018}
\acmYear{2018}
\acmDOI{XXXXXXX.XXXXXXX}

\acmConference[Conference acronym 'XX]{Make sure to enter the correct
  conference title from your rights confirmation emai}{June 03--05,
  2018}{Woodstock, NY}
\acmPrice{15.00}
\acmISBN{978-1-4503-XXXX-X/18/06}

\usepackage{bbding,tikz}
\usepackage{subcaption,threeparttable}
\usepackage{microtype}
\newcommand*\circled[1]{\tikz[baseline=(char.base)]{
    \node[shape=circle,draw,inner sep=0.5pt] (char) {#1};}}
    
\renewcommand\footnotetextcopyrightpermission[1]{}
\begin{document}

\title{High-performance and Scalable Software-based NVMe Virtualization Mechanism with I/O Queues Passthrough}



\author{Yiquan Chen}
\affiliation{%
  \institution{Zhejiang University }
  \institution{Alibaba Group}
  \country{China}
  }

\author{Zhen Jin}
\affiliation{%
  \institution{Zhejiang University}
  \country{China}
  }

  \author{Yijing Wang}
\affiliation{%
  \institution{Alibaba Group}
  \country{China}
  }

  \author{Yi Chen}
\affiliation{%
  \institution{Zhejiang University }
  \country{China}
  }

  \author{Hao Yu}
\affiliation{%
  \institution{Alibaba Group}
  \country{China}
  }

  \author{Jiexiong Xu}
\affiliation{%
  \institution{Zhejiang University}
  \country{China}
  }

  \author{Jinlong Chen}
\affiliation{%
  \institution{Zhejiang University }
  \country{China}
  }

  \author{Wenhai Lin}
\affiliation{%
  \institution{Zhejiang University }
  \country{China}
  }

  \author{Kanghua Fang}
\affiliation{%
  \institution{Alibaba Group}
  \country{China}
  }

  \author{Chengkun Wei}
\affiliation{%
  \institution{Zhejiang University }
  \country{China}
  }

  \author{Qiang Liu}
\affiliation{%
  \institution{Alibaba Group}
  \country{China}
  }

  \author{Yuan Xie}
\affiliation{%
  \institution{Alibaba DAMO Academy}
  \country{United States}
  }

  \author{Wenzhi Chen}
\affiliation{%
  \institution{Zhejiang University}
  \country{China}
  }








\renewcommand{\shortauthors}{sTrovato and Tobin, et al.}

\begin{abstract}
NVMe(Non-Volatile Memory Express) is an industry standard for solid-state drives (SSDs) that has been widely adopted in data centers. NVMe virtualization is crucial in cloud computing as it allows for virtualized NVMe devices to be used by virtual machines (VMs), thereby improving the utilization of storage resources. However, traditional software-based solutions have flexibility benefits but often come at the cost of performance degradation or high CPU overhead. On the other hand, hardware-assisted solutions offer high performance and low CPU usage, but their adoption is often limited by the need for special hardware support or the requirement for new hardware development.
 
In this paper, we propose LightIOV, a novel software-based NVMe virtualization mechanism that achieves high performance and scalability without consuming valuable CPU resources and without requiring special hardware support. LightIOV can support thousands of VMs on each server. The key idea behind LightIOV is NVMe hardware I/O queues passthrough, which enables VMs to directly access I/O queues of NVMe devices, thus eliminating virtualization overhead and providing near-native performance. Results from our experiments show that LightIOV can provide comparable performance to VFIO, with an IOPS of 97.6\%-100.2\% of VFIO. Furthermore, in high-density VMs environments, LightIOV achieves 31.4\% lower latency than SPDK-Vhost when running 200 VMs, and an improvement of 27.1\% in OPS performance in real-world applications. 
\end{abstract}





\maketitle

\begin{table*}[h!]
    \setlength{\abovecaptionskip}{0.2cm} 
    \setlength{\belowcaptionskip}{0.3cm}
    \centering
    \caption{Comparison of the existing NVMe virtualization and LightIOV}
    \label{table:compare}
    
    \begin{threeparttable}
    
    \setlength{\tabcolsep}{3pt} 
    \renewcommand{\arraystretch}{1.0} 
    \begin{tabular}{c c c c c c c c c c c}
    \toprule
      &\multicolumn{5}{c}{Hardware-assisted} &\multicolumn{4}{c}{Software-based} & \\
      \cmidrule(lr){2-6} \cmidrule(lr){7-10}
    \noalign{\smallskip}
   &VFIO &SR-IOV & SIOV & LeapIO &FVM &Virtio & SPDK-Vhost  & Mdev-NVMe & \textit{\textbf{LightIOV}}\\

    &\cite{williamson2012vfio} &\cite{sriov} & \cite{siov}&\cite{leapio}&\cite{kwon_fvm_nodate} & \cite{virtio} & \cite{spdk_2018}  & \cite{mdevnvme} &\\
        \noalign{\smallskip}
  \hline
      \noalign{\smallskip}
  High performance& \Checkmark&\Checkmark&\Checkmark & &\Checkmark & &\Checkmark&\Checkmark &\CheckmarkBold \\
    Low overhead& \Checkmark&\Checkmark&\Checkmark &\Checkmark & \Checkmark& &&&\CheckmarkBold \\
    Flexibility\tnote{*} &\Checkmark && & & &\Checkmark &\Checkmark  &\Checkmark &\CheckmarkBold \\
    
       High scalability&& \Checkmark &\Checkmark & \Checkmark & \Checkmark  & \Checkmark & \Checkmark & \Checkmark &\CheckmarkBold \\
   
    \bottomrule 
    \end{tabular}

    \begin{tablenotes}
        \footnotesize
        \item[*] Independent of special hardware support.
    \end{tablenotes}
    \end{threeparttable}
\end{table*}

\section{Introduction}

Storage virtualization is a key aspect of data centers as it allows for the provision of storage resources for virtual machines, decoupling virtual devices from physical devices, and allowing for flexible mapping of multiple virtual devices to a single physical device \cite{rosenblum2011virtualization}. In this way, storage virtualization can improve resource utilization and provide simple and consistent interfaces for complex functions \cite{keeriyadath2016nvme}, such as live migration \cite{kuperman2016paravirtual} and consolidation \cite{verma2010srcmap}.

As technology evolves, storage devices are also evolving to meet the performance demands of data centers. 
Solid state drives (SSDs) offer superior performance in terms of throughput and latency when compared to hard disk drives (HDDs) \cite{awad2015non, ananthanarayanan2011disk, narayanan2009migrating}. 
In addition, the Non-Volatile Memory Express  (NVMe) \cite{nvme} interface significantly improves the I/O performance of SSDs over the traditional SATA interface \cite{xu2015performance}. 
Thus, the NVMe interface has become an industry standard for SSDs, and NVMe SSDs are widely deployed in data centers such as AWS \cite{AWS}, Alibaba Cloud \cite{alibabacloud}, Microsoft Azure \cite{microsoft}, and Google Cloud \cite{GoogleCloud}. 


In data centers, NVMe virtualization mechanisms are used by cloud vendors to enable the virtualization of NVMe devices for multiple VMs. These mechanisms should be \textbf{high-performance}, \textbf{low-overhead}, and \textbf{flexible}. The primary objective of NVMe virtualization mechanisms is to minimize the performance degradation caused by virtualization, ensuring that performance is as close to bare metal as possible. Secondly, these mechanisms should operate with low overhead and consume minimal server resources (e.g., CPU cores). Finally, flexibility is an essential factor to consider. Avoiding the need for additional hardware to virtualize NVMe devices is optimal. The use of specialized NVMe SSDs (e.g., SR-IOV) or extra hardware for virtualization decreases flexibility for cloud vendors.

\textbf{Scalability} is an increasingly vital aspect as the density of VMs deployed on a single server continues to grow. Currently, it is common to  deploy hundreds of VMs on a single server \cite{zhang2019fast}, and lightweight VMs are widely used for serverless computing. With two-socket servers in the market capable of supporting up to 384 CPU cores/HTs \cite{AMD}, the VM density of a single server can easily reach into the thousands due to its massive computing capabilities. Moreover, the trend toward serverless computing is growing. Lightweight VMs \cite{agache2020firecracker,manco2017my,li2022rund}, which consume fewer CPU resources and memory than full-featured VMs, are commonly used in serverless for security and isolation reasons. In existing cloud data centers, there are already thousands of lightweight VMs deployed on each server \cite{agache2020firecracker,li2022rund}. Therefore, it is crucial that the NVMe virtualization mechanism has high scalability to support up to thousands of VMs.

\textbf{Limitations of the existing methods.} NVMe virtualization mechanisms are expected to be high-performance, low-overhead, flexible, and highly scalable. However, as shown in Table \ref{table:compare}, existing software-based and hardware-assisted solutions have several limitations in these dimensions.

\begin{itemize}
\item \textit{Software-based solutions} (e.g., virtio \cite{virtio}, SPDK vhost-NVMe \cite{spdk_2018}, Mdev-NVMe \cite{mdevnvme}) virtualize NVMe devices with pure software approaches. Therefore, they can work with general NVMe devices and show high flexibility. However, existing software-based solutions suffer from serious performance degradation or high CPU overhead. Virtio \cite{virtio} is an early approach that creates virtual device interfaces between guest OSes and hypervisors. But it significantly hampers the high performance of NVMe SSDs, with throughput only reaching 50\% of native performance \cite{mdevnvme}.
Then, polling-based solutions(e.g., SPDK vhost-NVMe \cite{spdk_2018}, MDev-NVMe \cite{mdevnvme}) were proposed. They use dedicated CPU cores for virtual I/O \cite{VirtualizationPollingEngine, Re-architectingVMMs,gavrilovska2007high} to improve the I/O performance of virtualized NVMe SSDs. However, the consumed host CPU cores are very valuable for cloud vendors.
For example, in our cloud instances, we use the SPDK vhost-NVMe solution. A typical server with 12 SSDs requires allocating 8-10 CPU cores to perform storage virtualization. Moreover, as the VM density increases, it requires even more CPU resources to maintain high performance.

\item \textit{Hardware-assisted solutions} (e.g., VFIO \cite{vfiouser}, SR-IOV \cite{sriov}, SIOV \cite{siov}, FVM \cite{kwon_fvm_nodate}, and LeapIO \cite{leapio}) virtualize NVMe devices at the hardware level, providing high performance and no CPU usage. However, these solutions lack flexibility or require the development of additional dedicated hardware.
In particular, while NVMe devices with SR-IOV or SIOV capability can directly virtualize the device, this approach increases the complexity of the SSD controller and is only supported by a small number of NVMe SSDs. On the other hand, developing dedicated hardware for NVMe virtualization can be an option, such as FVM \cite{kwon_fvm_nodate} and LeapIO \cite{leapio}, but extra hardware leads to higher hardware costs, power consumption, and deployment difficulty.


\end{itemize}


In this paper, we propose LightIOV, a novel software-based NVMe virtualization mechanism that achieves high performance and high scalability without consuming valuable CPU resources and without requiring special hardware support. The key idea of LightIOV is NVMe I/O queues passthrough, which enables VMs directly access the I/O queues of NVMe devices. 
Specifically, LightIOV consists of three components: LightIOV frontend driver, LightIOV backend driver, and LightIOV device. The LightIOV frontend driver is responsible for presenting a standard NVMe device to the guest operating system, eliminating the need for any modifications to the guest's applications or other operating system components.
The LightIOV backend driver creates I/O queues and maps the I/O queues buffer to VMs, allowing VMs to access I/O queues directly. Additionally, the backend driver takes advantage of IOMMU support to enable DMA transactions and interrupt processing for guest VMs without host software involvement.
Finally, the LightIOV device in the hypervisor emulates control resources (PCIe configuration, BAR space, and admin queue) and combines them with data resources (NVMe I/O queues, doorbell registers, interrupt resources, and LBAs) allocated by the LightIOV backend driver.


We evaluated the LightIOV prototype and compared it with existing NVMe virtualization mechanisms, including hardware-assisted VFIO \cite{vfiouser}, and software-based SPDK vhost-NVMe \cite{spdk_2018}, virtio \cite{virtio}. 
The results demonstrate that LightIOV can provide near-native NVMe performance with an IOPS of 97.6\%-100.2\% of VFIO. Furthermore, in high-density VMs environments, LightIOV achieves 31.4\% lower latency than SPDK-Vhost when running 200 VMs, and an improvement of 27.1\% in OPS performance in real-world applications.

In summary, our contributions are as follows:

\begin{itemize}
\item
 We propose LightIOV, a novel software-based NVMe virtualization mechanism achieving near-native performance and high scalability while not consuming valuable CPU resources and not requiring special hardware support.
 
\item
We design and implement NVMe I/O queues passthrough, which enables VMs directly access the I/O queues of NVMe devices.

 \item
 We introduce a new highly scalable NVMe virtualization mechanism that emulates full NVMe devices for thousands of VMs through software.

 \item
We implement LightIOV and conduct ample comparison evaluations between LightIOV and other virtualization solutions in terms of I/O performance, scalability, and fairness. Both synthetic benchmarks and real-world application results demonstrate the superiority of LightIOV over other solutions.
\end{itemize} 

\begin{figure}[t] 
   \centering
    \begin{minipage}[t]{\linewidth}
        \centering
        \setkeys{Gin}{width=\linewidth}  \includegraphics{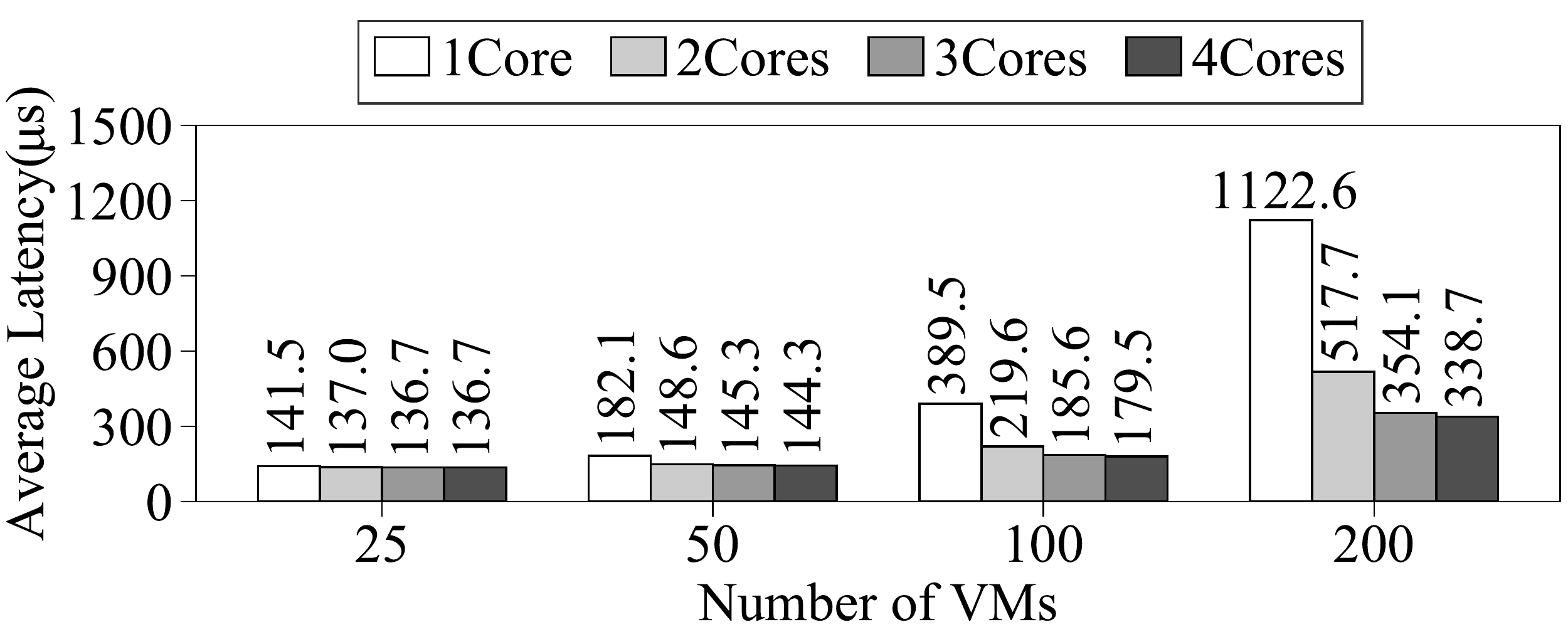}
    \end{minipage}
    \caption{Average Latency of SPDK vhost-NVMe with different CPU cores for polling. As the VM density increases,  SPDK vhost-NVMe requires more valuable CPU resources to maintain high performance. }
     \label{fig:motivation}
\end{figure}
\section{Background}

This section gives background on NVMe and virtualization solutions.

\subsection{NVM Express}
Non-Volatile Memory (NVM) technologies boost the inexorable trends to create ever more powerful storage devices. However, due to the limitation of the Advance Host Controller Interface (AHCI) \cite{ahci_sata} architecture design in Serial ATA (SATA ) \cite{ahci_sata} interface, high-performance storage devices remain seriously underutilized. Therefore, to settle with the I/O performance bottleneck, NVM Express (NVMe) \cite{nvme} is proposed to exploit the potential of fast NVM storage and optimize I/O paths with a large number of I/O queues depth. 

The NVMe specification allows for one Admin Queue Pair to be associated with up to \textbf{65,535 I/O Queue Pairs}, with the Admin Queue Pair responsible for executing functions that impact the entire controller, such as creating I/O Queue Pairs, namespace management, and setting features. The I/O Queue Pair, on the other hand, is used specifically for carrying out I/O operations (read/write). Each Queue Pair (QP) comprises a Submission Queue (SQ) and a Completion Queue (CQ), with the former responsible for submitting commands and the latter for receiving completions for those commands. Both SQ and CQ are circular buffers in host memory that are shared with the device through Direct Memory Access (DMA). Each queue (SQ or CQ) maintains a record of the head or tail pointer of the circular buffer in its Doorbell Registers(DBs).
\textbf{Controller Memory Buffer (CMB)} is a general-purpose read/write memory region on the NVMe device controller that can be  utilized for various purposes. For example, storing SQs in CMB enables the NVMe driver to write the entire SQ entry directly to the controller's internal memory space, avoiding fetching the entry from host memory.

The processing of the NVMe command is as follows: the NVMe driver places NVMe commands in the SQ and rings the target SQ doorbell register to indicate that new commands are generated. Then, the NVMe device fetches and processes the newly added commands. Later, the NVMe device reads/writes data from/to the host via DMA. Once the NVMe commands are completed, the NVMe device writes completion messages to the associated CQ and generates an interrupt. Finally, the NVMe driver handles the CQ entries and updates the target CQ doorbell register to clear the interrupt and release the CQ entries.

\subsection{NVMe Virtualization} \label{StorageVirtualization}

\emph{1) Software-based virtualization.} Software-based storage virtualization mechanisms can be divided into three categories: full-virtualization, para-virtualization, and polling-based virtualization.

In full-virtualization, the hypervisor emulates the functionality of the hardware device according to the hardware specification with software. It is transparent to the VMs and does not require special hardware support. However, it has faded out of view due to suffering from severe performance degradation. 

Para-virtualization emerged as a solution to address the performance issues of full-virtualization, and virtio \cite{virtio} is a de-facto standard for para-virtualized driver specification. Virtio drivers are composed of frontend and backend parts, i.e., virtio frontend drivers in the guest OS kernel and virtio backend drivers in the hypervisor. The two parts negotiate with each other on virtual I/O queues via shared memory. 

Due to inefficient virtio backend drivers, powerful NVMe devices cannot be fully utilized \cite{spdk_2018}. Polling-based approaches \cite{mdevnvme, SPDK_2017,spdk_2018} use dedicated CPU cores to achieve high performance. Among all SPDK vhost-target solutions \cite{spdkvhost} (including SPDK vhost-scsi, SPDK vhost-blk, SPDK vhost-NVMe), SPDK Vhost-NVMe has the best performance for NVMe devices with optimization for NVMe SSDs \cite{spdk_2018}.
The SPDK vhost-NVMe solution works as follows: (1) A thread is run on the dedicated CPU core to poll the virtual NVMe I/O queues in the shared memory area. (2) The new request is converted into an NVMe command. (3) The request is then sent to the physical device via an SPDK user-level NVMe driver. (4) NVMe device reads/writes data from/to the guest VM. (5) After the request completes, the thread handles the request completion information and injects the interrupt into the VM via the hypervisor. 

 Similarly, Mdev-NVMe \cite{mdevnvme} makes use of a mediated pass\-through mechanism and designs active polling for shadowed SQs and CQs and host CQs in the kernel to gain high performance.

\emph{2) Hardware-assisted virtualization.}
The direct device assignment (or VFIO) enables VMs directly use NVMe devices. It makes use of I/O Memory Management Units (IOMMU) support for DMA and interrupts remapping (e.g., Intel-VTd \cite{intel_vt}, AMD-Vi \cite{amd_iommu}), which eliminates the overhead associated with software virtualization and leads to near-native performance. However, VFIO requires that physical devices be exclusively assigned to a single VM, thereby sacrificing the resource-sharing feature of virtualization.

To address the lack of shareability in VFIO, PCIe SR-IOV \cite{sriov} is proposed, which enables sharing of a physical device among multiple VMs at the hardware level. SR-IOV-capable storage devices consist of a physical function (PF) and multiple virtual functions (VFs). The PFs are managed by the host software and are responsible for the resource management of the device, while each VFs is attached to a PF and can be assigned to a VM. SR-IOV-capable devices can achieve both near-native disk performance and device sharing. FVM \cite{kwon_fvm_nodate}, a hardware-assisted storage virtualization mechanism, which implements the SR-IOV layer on an FPGA card to achieve high performance and scalability while supporting various VM features. Besides, LeapIO \cite{leapio} offloads the entire storage stack to the ARM SoC to reduce the burden on the host CPU.


Scalable IOV (SIOV) \cite{siov} is a hardware-assisted I/O virtualization specification focusing on the efficient and scalable sharing of I/O devices. Specifically, SIOV allows more frequent and performance-critical operations to be run directly on hardware while complex control and configuration operations are emulated through software. 

\section{Motivation}
\label{motivation}

This section discusses the challenges of existing NVMe virtualization mechanisms in data centers. We have observed that software-based solutions suffer from either performance degradation or high CPU overhead. Meanwhile, hardware-assisted solutions encounter the issue of poor flexibility that requires special hardware support.

\subsection{Disadvantages of existing software-based virtualization}

{\bf Performance degradation.} Full-virtualization and para-virtualization both suffer from significant performance degradation. For example, virtio, a traditional para-virtualization solution, can only achieve 50\% of native performance \cite{mdevnvme}.
This degradation can be attributed to two factors. First, the costly VM\_Exit events are generated when VMs dispatch requests to SQs, and the hypervisor generates virtual interrupts to VMs. These events cause the guest to be suspended and the host to be resumed, resulting in performance degradation due to context switching overhead and cache pollutions \cite{forvirtio,TurtlesProject, ComparisonofSoftwareandHardware, SplitX}. Second, for I/O operations between the NVMe devices and VMs, the hypervisor needs to perform at least two data movements, which can reduce I/O performance \cite{gordon_eli_2012}.

{\bf High CPU overhead.}  Despite providing near-native performance, advanced polling-based NVMe virtualization (e.g., SPDK vhost-target \cite{spdk_2018}, Mdev-NVMe \cite{mdevnvme}) incurs high CPU overhead. In our testing of the SPDK vhost-NVMe solution, we found that it requires 1 CPU core to reach the IOPS limit of 2 Intel P4510 SSDs. In addition, cloud vendors often offer storage services with features such as hot upgrades and live migration, which increase software complexity and demand more CPU resources. As a result, virtualizing NVMe storage on a server with 12 SSDs can typically consume 8-10 CPU cores in a production environment. Since servers are commonly configured with 96-128 CPU cores, storage virtualization overhead can amount to as much as 10.4\% of CPU resources.
 
Besides, polling-based solutions waste valuable CPU resources during periods when VMs are not generating I/O workloads. In polling-based solutions, dedicated CPU cores must continuously poll relevant memory regions to process virtual I/O requests. As a result, 100\% of the CPU cycles are consumed, even when no new I/O requests are present. This inefficient usage of CPU resources can lead to reduced system performance and increased energy consumption.

Furthermore, polling-based solutions have another inherent drawback: as VM density increases, these solutions need more CPU resources to maintain high performance.
Figure \ref{fig:motivation} illustrates the impact of assigning different numbers of CPU cores for SPDK-Vhost polling when running 25, 50, 100, and 200 VMs with the 4k-randread-1-1 workload on two SSDs. It shows that when running 25 VMs, there is no significant performance difference with different numbers of CPU cores used for polling. However, when running 200 VMs, the I/O latency of using two cores for polling is 53.9\% lower than using only one core. Therefore, as VM density increases, polling-based solutions require more valuable CPU resources to maintain their high-performance benefits. Otherwise, they can lead to significant performance degradation.

\subsection{Limitations of existing hardware-assisted virtualization}
{\bf Lack of flexibility.} Hardware-assisted mechanisms require special hardware features, which limits their flexibility. 
 NVMe SSDs with SR-IOV or SIOV capability directly virtualize these devices by SSD controllers. However, only a limited number of SSDs support the SR-IOV feature, and currently, no NVMe SSDs support SIOV. In addition, SIOV not only demands SSDs to have SIOV-capability but also requires the latest server CPUs with emerging capabilities (e.g., PASID-granular address translation by DMA remapping hardware). While it is possible to standardize these virtualization capabilities for all NVMe SSDs to support, it will increase the complexity of the SSD controller and cause compatibility issues with NVMe devices from different manufacturers.

{\bf TCO challenges of dedicated hardware.}
Using dedicated hardware is another approach for NVMe virtualization, such as FVM \cite{kwon_fvm_nodate}, and LeapIO \cite{leapio}. However, dedicated hardware has its own challenges in terms of the total cost of ownership (TCO). FVM has to develop new hardware based on FPGA, which will incur extra hardware costs and power consumption. LeapIO offloads the entire storage stack to the ARM SoC, and it supports both local and remote storage. However, it suffers from severe performance degradation, achieving only 68\% \cite{kwon_fvm_nodate} throughput of the single native disk due to the limited computing capabilities of the ARM CPU.


\subsection{Summary}
As summarized in Table \ref{table:compare}, existing software-based NVMe virtualization solutions suffer from significant performance degradation or high CPU overhead, while hardware-assisted solutions lack flexibility and have TCO challenges. These challenges have motivated us to propose LightIOV, which eliminates virtualization overhead, achieves near-native performance, and does not require special hardware support.
 
\begin{figure}
  \centering
  \setlength{\abovecaptionskip}{-0.1cm}
  \setkeys{Gin}{width=\linewidth}
\includegraphics{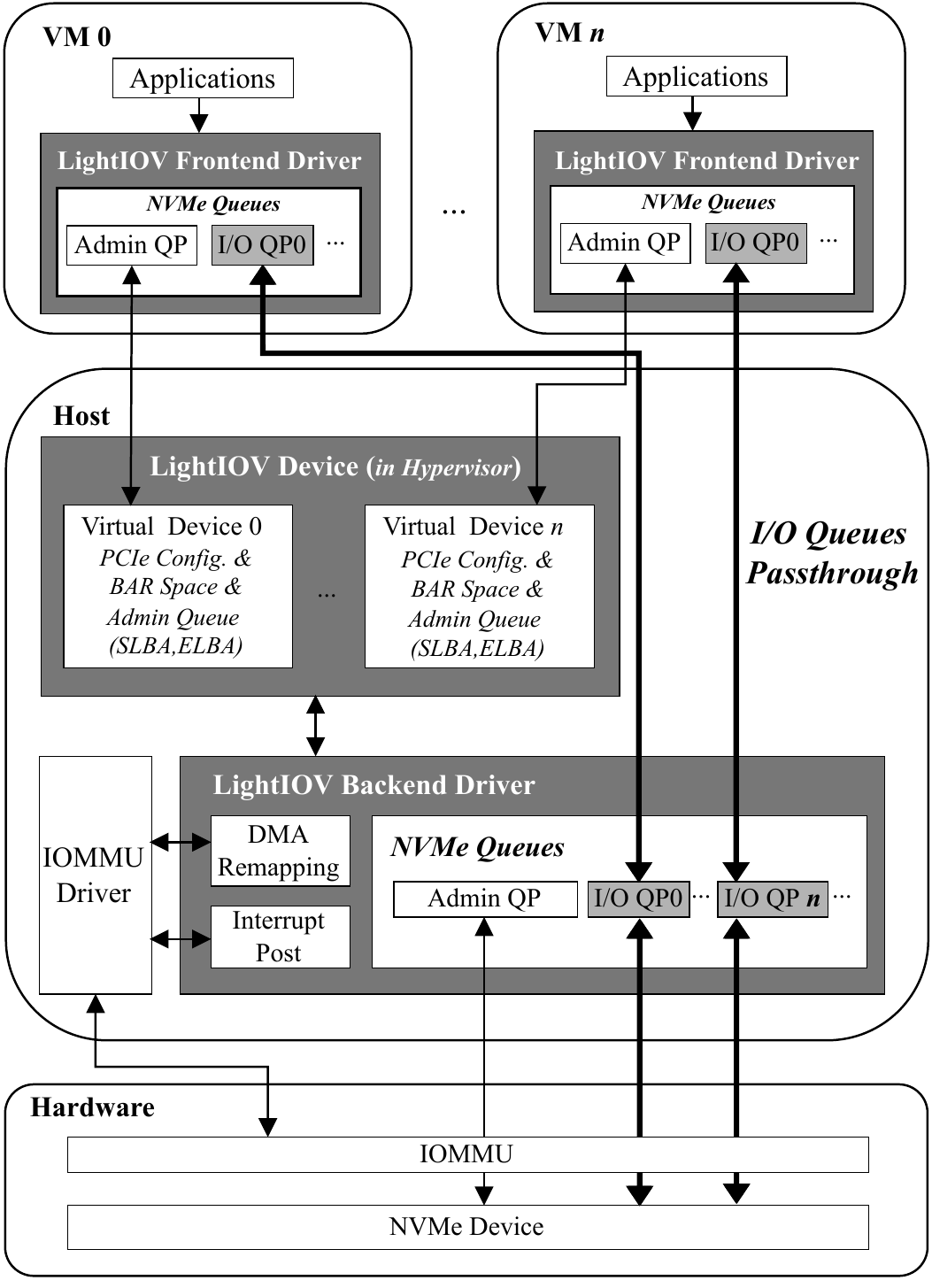}
  \hspace{0.08\linewidth}
  \caption{Overall Architecture of LightIOV}
  \label{fig:Overall Architecture of LightIOV}
\vspace{-0.3cm}
\end{figure}
\section{Design and Implementation}

In this section, we introduce the design of LightIOV and its implementation. 
\subsection{Design Goals}
 We reviewed state-of-the-art solutions in NVMe virtualization and found that existing software-based and hardware-assisted solutions have different advantages and disadvantages. Software-based solutions offer high flexibility by being independent of hardware, while hardware-assisted solutions virtualize NVMe devices at the hardware level, allowing VMs to interact efficiently with physical devices without consuming the host's CPU resources. To this end, we set the following design goals for LightIOV:

 {\bf High performance.} The mechanism should provide performance in terms of throughput and latency that is comparable to that of native NVMe SSDs.
 
 {\bf Low overhead.} It should not consume valuable CPU resources, as CPU cores in cloud computing environments are limited and valuable. More applications can be run by saving CPU resources, making it an attractive option for cloud vendors looking to reduce TCO.

 {\bf Flexibility.} The mechanism should be applicable to all NVMe devices, regardless of whether they have virtualization capabilities such as SR-IOV, and should not require additional dedicated hardware for virtualization.
 
 {\bf High scalability.} The NVMe virtualization mechanism should be able to scale nicely with the increasing density of VMs on a single server, as the density of VMs is continuously increasing and lightweight VMs such as Firecracker \cite{firecracker} and RunD \cite{rund} are being widely deployed in cloud data centers. It should be able to support thousands of VMs and maintain its high performance.
  
\subsection{Overall Architecture}
To meet the aforementioned design goals, we propose LightIOV, a novel software-based NVMe virtualization mechanism that has the advantages of software-based solutions (flexibility) and hardware-assisted solutions (high performance and low CPU overhead).

 LightIOV divides the NVMe device resources into \textbf{control resources} (PCIe configuration, BAR space, and admin queue) and \textbf{data resources} (I/O queues, doorbell registers, interrupt resources, and LBAs). Because the NVMe device has only one control resource, but each emulated device must have its own separate control resources, LightIOV must virtualize control resources in the hypervisor for sharing. Control resources are utilized for device initialization, querying, and management and are not involved in I/O requests. Therefore, using software emulation for control resources does not have any impact on I/O performance. An NVMe device can accommodate up to 65,535 I/O queues according to NVMe specifications \cite{nvme}, so LightIOV does not need to virtualize these resources and I/O queues can be directly assigned to different VMs.

Figure \ref{fig:Overall Architecture of LightIOV} shows the LightIOV architecture and its components. LightIOV consists of three components: LightIOV backend driver in the host kernel,  LightIOV frontend driver in guest VMs, and LightIOV device in the hypervisor.

 \textbf{LightIOV backend driver} is responsible for enabling VMs capable of directly accessing I/O queues, DMA remapping, and Interrupt Post. It creates I/O queues and maps the I/O queues buffer to VMs via the virtual device's Control Memory Buffer (CMB), then VMs can access I/O queues directly. DMA Remapping and Interrupt Post components achieve DMA transactions and interrupt processing for guest VMs without hypervisor involvement.

 \textbf{LightIOV device} virtualizes NVMe control resources and emulates full NVMe devices for VMs. Each virtual NVMe device has control resources (PCIe configuration, BAR space, and admin queue) and data resources (I/O queues, doorbell registers, interrupt resources, and LBAs). The data resources are allocated by the LightIOV backend driver. LightIOV Device manages virtual-NVMe Logical Block Addressing(LBA) resources.

 \textbf{LightIOV frontend driver} provides standard NVMe devices for guest VMs, so user applications or other operating system components need not do any modifications. The LightIOV frontend driver directly leverages the I/O queues in the CMB provided by LightIOV Device for creating NVMe I/O queues. Because of NVMe I/O queues passthrough, the LightIOV frontend driver is also responsible for converting the virtual LBA to the physical LBA of the NVMe device only by adding an offset.
 
\begin{figure}
  \centering
  \setlength{\abovecaptionskip}{-0.cm}
  \setkeys{Gin}{width=\linewidth}
    \includegraphics{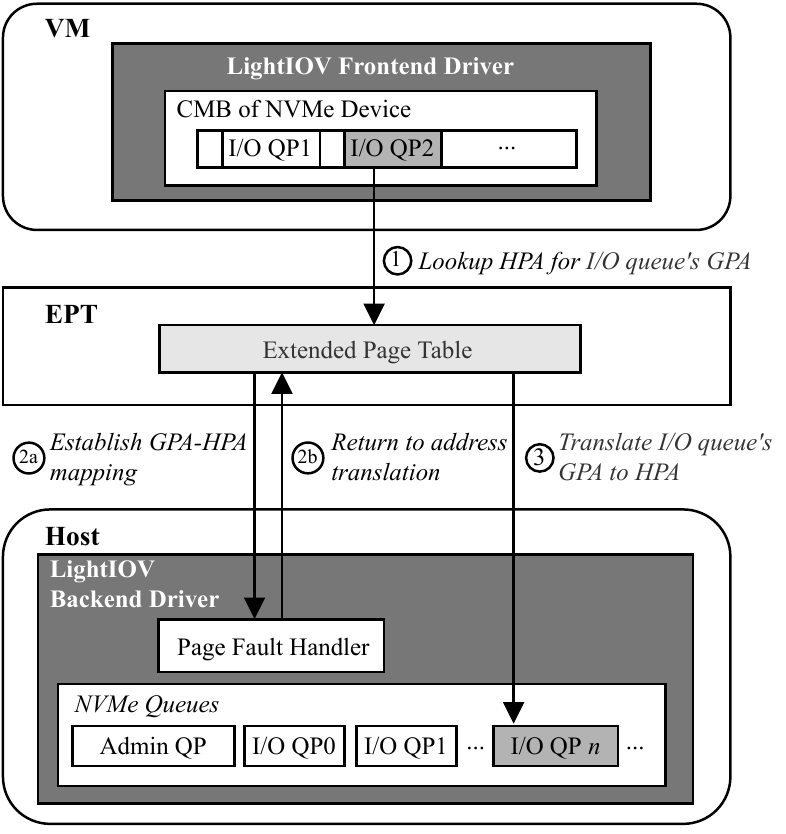}
  \hspace{0.08\linewidth}
  \caption{I/O Queues Passthrough Mechanism}
  \label{fig:Process of Guest I/O Commands}
  \vspace{-0.4cm}
\end{figure}

\subsection{How I/O Queues passthrough works}
The Controller Memory Buffer(CMB) of NVMe is a general-purpose memory inside the NVMe devices that can be used for various purposes. The principle of I/O queues passthrough is that the LightIOV backend driver uses CMB to map the I/O queues memory between the host and guest VMs, which is achieved by the registered page fault handler to establish the guest I/O queues GPA-HPA mapping in Extended Page Table (EPT). In this way, VMs can write I/O requests to host NVMe I/O queues directly with the EPT address translation. 
We show how I/O queues passthrough works in Figure \ref{fig:Process of Guest I/O Commands}.

\circled{1} The LightIOV frontend driver in VM submits I/O requests to the I/O SQ in the CMB, and the Memory Management Unit (MMU) converts the Guest Virtual Address (GVA) of the I/O queue to Guest Physical Address (GPA). Then, the EPT takes over the GPA from the MMU and lookup HPA for the I/O queues GPA in the EPT.
If the GPA is not mapped to any host memory, it will generate a page fault interrupt, then go to \circled{2}; otherwise, go to \circled{3}.

\circled{2} LightIOV implements its own page fault handler. The page fault handler obtains the HPA of the corresponding I/O QPs and establishes the GPA-HPA mapping in EPT (\circled{2a}). After the address mapping is completed, it returns to HPA-GPA translation in EPT (\circled{2b}). 
The GPA-HPA mappings of doorbell registers are established similarly. The page fault handler is only called once for each I/O queue. Once the I/O queue page table is established, I/O requests on this queue will not cause a page fault, so it has no impact on I/O performance.

\circled{3} The EPT translates the guest I/O queues GPA to HPA according to the established page table entry. So the VM can access I/O SQ on the host memory to submit the I/O command, which is then fetched and executed by the NVMe device.

Consequently, once page table entries for I/O queues and doorbell registers are established in the EPT, virtual machines can access NVMe I/O queues directly, just like the host. Given that NVMe SSDs are capable of sustaining several hundreds of thousands of I/O operations per second, the overhead associated with page fault events triggered by one of these I/O requests is negligible. Additionally, EPT is generally available on existing servers, and LightIOV ingeniously employs this existing address translation mechanism to enable I/O queues passthrough.


\begin{figure}
  \centering
  \setlength{\abovecaptionskip}{0.3cm}
  \setkeys{Gin}{width=\linewidth}
    \includegraphics[width=0.4\textwidth]{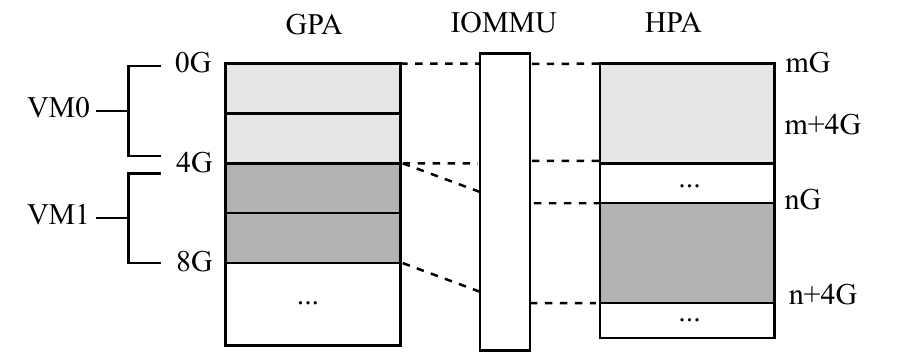}
  \hspace{0.08\linewidth}
  \caption{ DMA address isolation in IOMMU}
  \label{fig:non-overlapping GPA addresses}
\vspace{0cm}
\end{figure}

\subsection{DMA Remapping \& Interrupt Post} \label{dmaremapping}
DMA data transfer and interrupt handling are essential for a high-performance NVMe virtualization mechanism. With IOMMU support for DMA and interrupt remapping, LightIOV enables VMs to transfer data with the NVMe device by DMA and process interrupts for VMs without going through costly VM\_Exit.

{\bf DMA Remapping.} The DMA address in the guest I/O request is a Guest Physical Address (GPA). The GPA must be converted to a Host Physical Address (HPA) for NVMe devices initiating DMA transactions, and the IOMMU performs GPA-HPA address translation automatically. 
However, since the IOMMU uses the BDF (Bus, Device, Function) of PCIe devices to distinguish page table entries, different VMs accessing the same PCIe device share the IOMMU page table entry too. 
The GPA from different VMs may be the same, so may be translated to the same HPA, resulting in incorrect DMA transactions due to GPA conflicts.
To guarantee the DMA address isolation of different VMs, the hypervisor sets non-overlapping GPA addresses for VMs at boot time. For example, as shown in Figure \ref{fig:non-overlapping GPA addresses}, the hypervisor allocates GPA from 0G to 4G for VM0 and 4G to 8G for VM1. Therefore, the DMA addresses in the I/O commands of VM0 and VM1 are mapped to different HPA after IOMMU translation.

{\bf Interrupt Post.} LightIOV leverages IOMMU interrupt remapping unit and combines it with the post-interrupt, which is a hardware mechanism that allows interrupts to be received directly by a VM. LightIOV backend driver configures the post-interrupt I/O queues with the \textit{IRQ bypass manager}  \cite{irqbypass}.

In this way, when the interrupt of the I/O request reaches the IOMMU, the IOMMU hardware queries the relevant Interrupt Remapping Table Entry and translates the physical interrupt into a virtual interrupt for VMs. Then the virtual interrupt is injected into the VM and processed by the interrupt handler registered by the VM. The whole interrupt processing does not require the involvement of the host software or the exit of VMs, which can effectively improve the efficiency of the VMs handling interrupts, thus improving I/O performance.

\begin{figure}
  \centering
  \setlength{\abovecaptionskip}{-0.cm}
  \setkeys{Gin}{width=\linewidth}
    \includegraphics{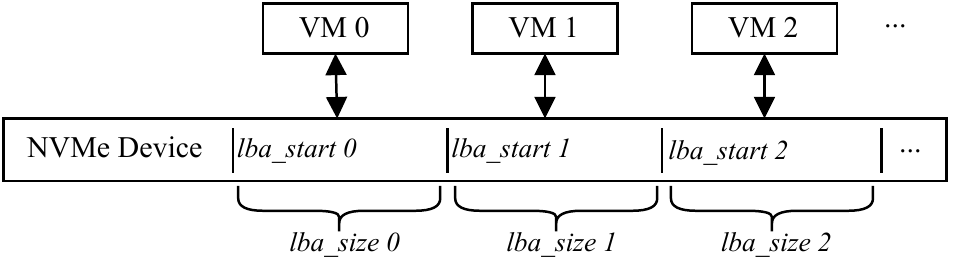}
  \caption{Isolation of NVMe device storage by assigning different LBA ranges to VMs}
  \label{fig:labrange}
  \vspace{-0.1cm}
\end{figure}

\subsection{Virtual Device Emulation}\label{Virtual Device Abstration}

LightIOV presents full NVMe devices for VMs in the hypervisor, which emulates the control resources through software and combines them with data resources allocated by the LightIOV backend driver.
The control resources, including PCIe configuration, BAR space, and admin queue, are essential to an NVMe device. LightIOV uses a conventional trap-and-emulate approach to emulate control resources in the hypervisor.
When VM access to the control resources of the virtual device causes VM\_Exit events and traps to the host, and then the hypervisor takes over the VM request. Then, the hypervisor reads/updates the relevant virtual registers. Lastly, the hypervisor generates an interrupt to the VM to inform the request completion. 
In general, only admin commands for device initialization, management, and status queries access control resources. In contrast, I/O requests for reading and writing data, which are crucial for determining I/O performance, do not involve control resources. Therefore, using the trap-and-emulate approach for control resources does not affect I/O performance.


The LightIOV device combines control resources (PCIe configuration, BAR space, and admin queue) and data resources (I/O queues, doorbell registers, interrupt resources, and LBAs) to present full NVMe devices for VMs.
The number of virtualized NVMe devices that LightIOV can provide is limited by  the number of NVMe I/O Queues implemented in SSD. 

According to NVMe specifications \cite{nvme}, the maximum number of I/O queues of one NVMe device is 65,535, so LightIOV can easily scale up to \textbf{tens of thousands VMs}.

  \begin{table}
  \caption{Test Cases}
  \label{table:test cases}
  \begin{tabular}{ll}
    \toprule
    \textbf{Case name} & \textbf{(bs,rw,numjobs,iodepth)}\\
    \midrule

     randread-4k-4-128 & (4k,randread, 4,128) \\
      
      randwrite-4k-4-128 &  (4k,ranwrite, 4,128) \\
      
      seqread-128k-4-128 & (128k,read,4,128) \\
      
      seqwrite-128k-4-128 &  (128k,write,4,128) \\
      
      randread-4k-1-1 &  (4k,randread,1,1) \\
      
      randwrite-4k-1-1 &  (4k,randwrite,1,1) \\
      
     seqread-4k-1-1 &  (4k,read,1,1) \\
      
     seqwrite-4k-1-1 &  (4k,write,1,1) \\
  \bottomrule
\end{tabular}
\end{table}

\subsection{Isolation}
As VMs can send I/O requests directly to the physical device after queue mapping is complete, checking the validity of the destination address in I/O requests is essential to prevent unauthorized access between VMs.

\textbf{LBA isolation.} An NVMe device leverages Logical Block Address (LBA) to address its internal space.
To ensure that each VM can only access its designated disk area, LightIOV implements the LBA address isolation through the cooperation of the LightIOV frontend driver and backend driver. Specifically,  as illustrated in Figure \ref{fig:labrange}, the LightIOV backend driver allocates non-overlapping LBA address ranges for each VM and stores the corresponding \verb|lba_start| and \verb|lba_size| information in the CMB of the virtual device. When a VM submits I/O commands via \verb|nvme_submit_cmd|, the LightIOV frontend driver adds the \verb|lba_start| to the \verb|slba| of the I/O command, and checks if the resulting LBA is within the VM's accessible range. If it is outside the range, an I/O error occurs, and the request cannot be sent.

\textbf{GPA isolation.} 
Section \ref{dmaremapping} explains how the hypervisor assigns non-overlapping GPA addresses for each VM  to ensure that each VM accesses its own HPA after the IOMMU translation. To further prevent malicious VMs from reading and writing the memory of other VMs (for instance, if VM0 has a GPA range of 0-4G, but the malicious VM0 issues an address space with a GPA of 6G), the LightIOV frontend driver performs a legitimacy check on the target memory address (GPA) of I/O commands. Specifically, the LightIOV frontend driver verifies whether the \verb|PRP| address in the NVMe request sent by the guest is within the GPA address range of the VM. If it is not, LightIOV generates an I/O error, and the request cannot be sent to the NVMe device.


\begin{figure}[tbp]
\centering
\begin{minipage}{\linewidth}
\centering
   \setkeys{Gin}{width=\linewidth}   
   \setlength{\abovecaptionskip}{-1.3cm}
    \includegraphics{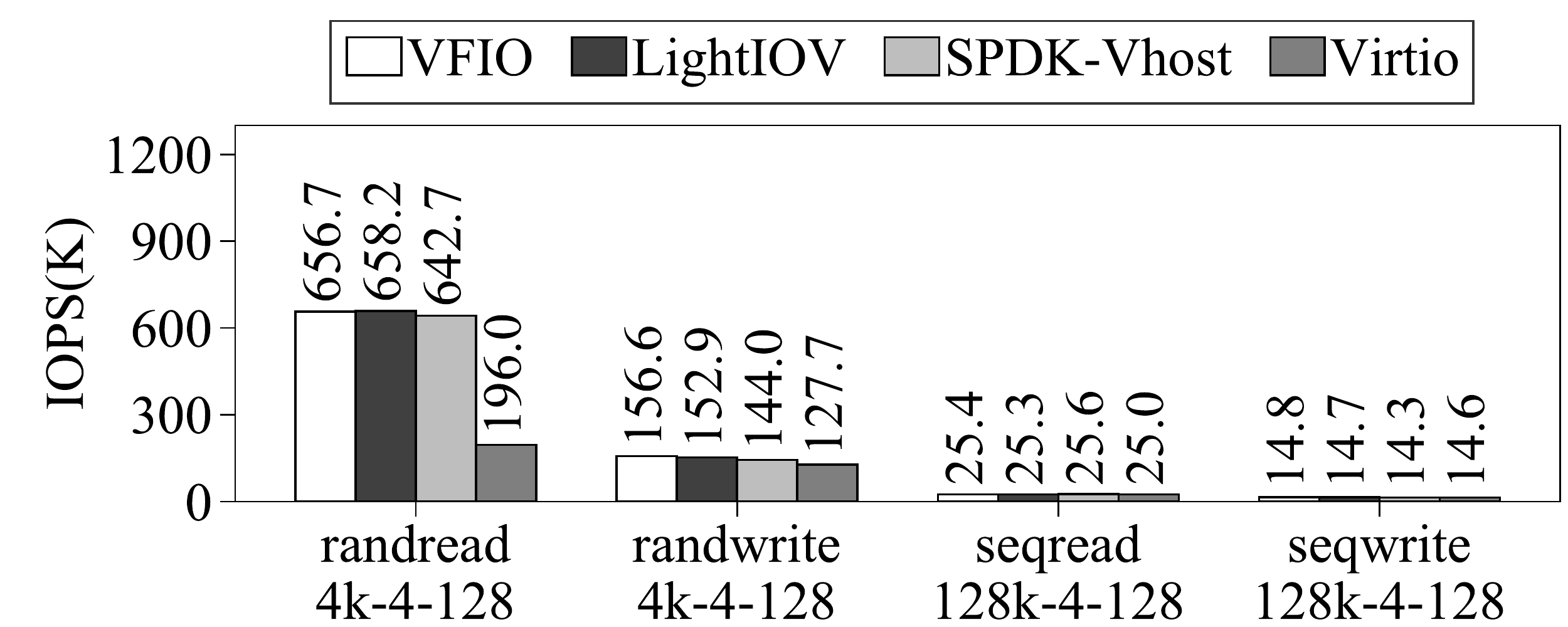}
  \subcaption{IOPS}
  \vspace{0.2cm}
  \label{fig:sig_IOPS}
\end{minipage}
\begin{minipage}{\linewidth}
  \centering
   \setkeys{Gin}{width=\linewidth}   
    \includegraphics{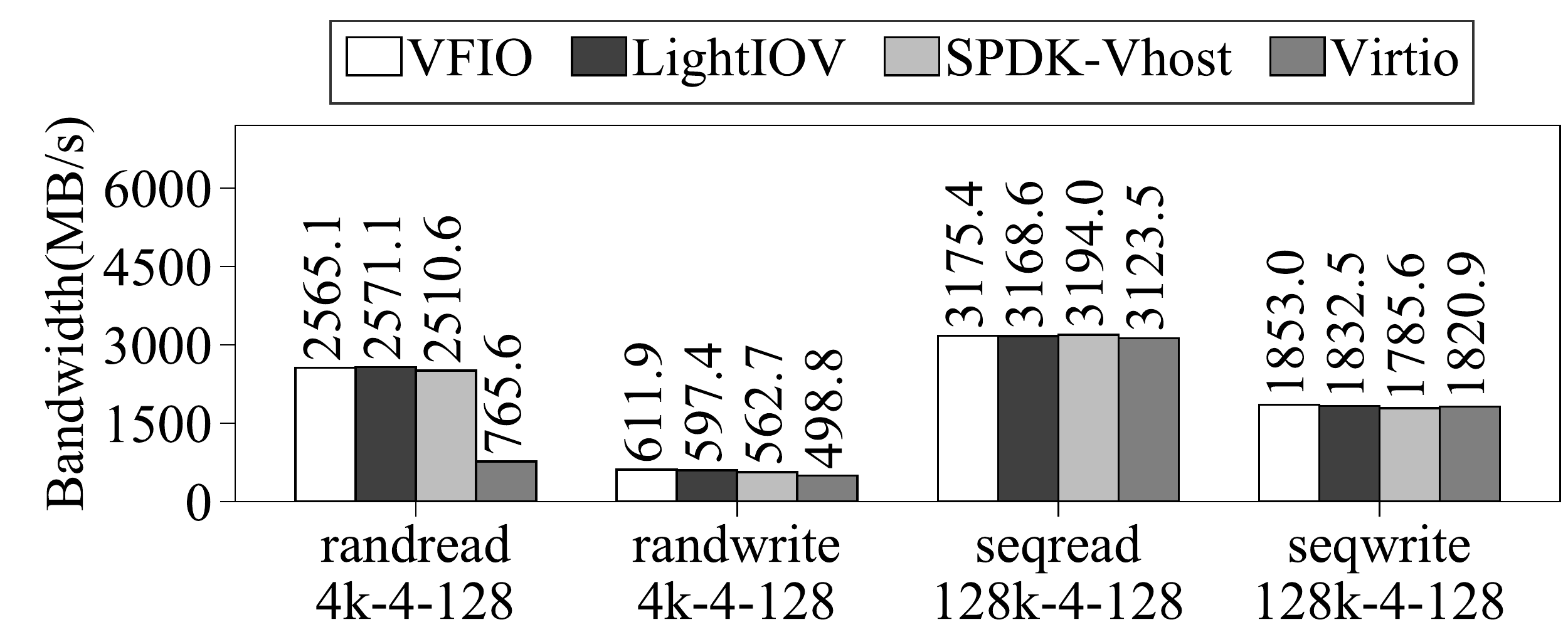}
  \subcaption{Bandwidth}
  \label{fig:sig_BW}
\end{minipage}
 
\caption{Single-VM throughput with one SSD}
 \label{fig:sig}
\vspace{-0.3cm}
\end{figure}

\subsection{Implementation}

We implemented LightIOV on the Linux kernel (version 4.19) and Firecracker (version 0.14.0). Firecracker \cite{firecracker} is a new open-source Virtual Machine Monitor (VMM), also called a hypervisor. LightIOV implementations contain 6,600 LOC in total, including 300 LOC in the guest kernel, 4,200 LOC in the host kernel, and 2,100 LOC in Firecracker.

For the guest kernel, we modified the original NVMe driver to the LightIOV frontend driver, which directly leverages the I/O queues in the CMB and adds an offset to the \verb|slba| of guest NVMe commands to realize the storage isolation between VMs.
In the host kernel, we implemented the LightIOV backend driver based on the original NVMe driver and the VFIO-mdev framework \cite{mdev_linux}. The LightIOV backend driver creates NVMe I/O queues on the host. Then it exposes these I/O queues along with the doorbell registers and LBA range to the hypervisor through the VFIO-mdev framework.
In Firecracker,  we added support for VFIO devices and adopted the VFIO-mdev framework \cite{mdev_linux} to provide NVMe \textsl{data resources} for LightIOV Device.
The emulations for PCIe configuration space, BAR space, and NVMe admin queue are also implemented in Firecracker.





\begin{figure}[tbp]
\centering
\begin{minipage}{\linewidth}
\centering
   \setkeys{Gin}{width=\linewidth}   
    \includegraphics{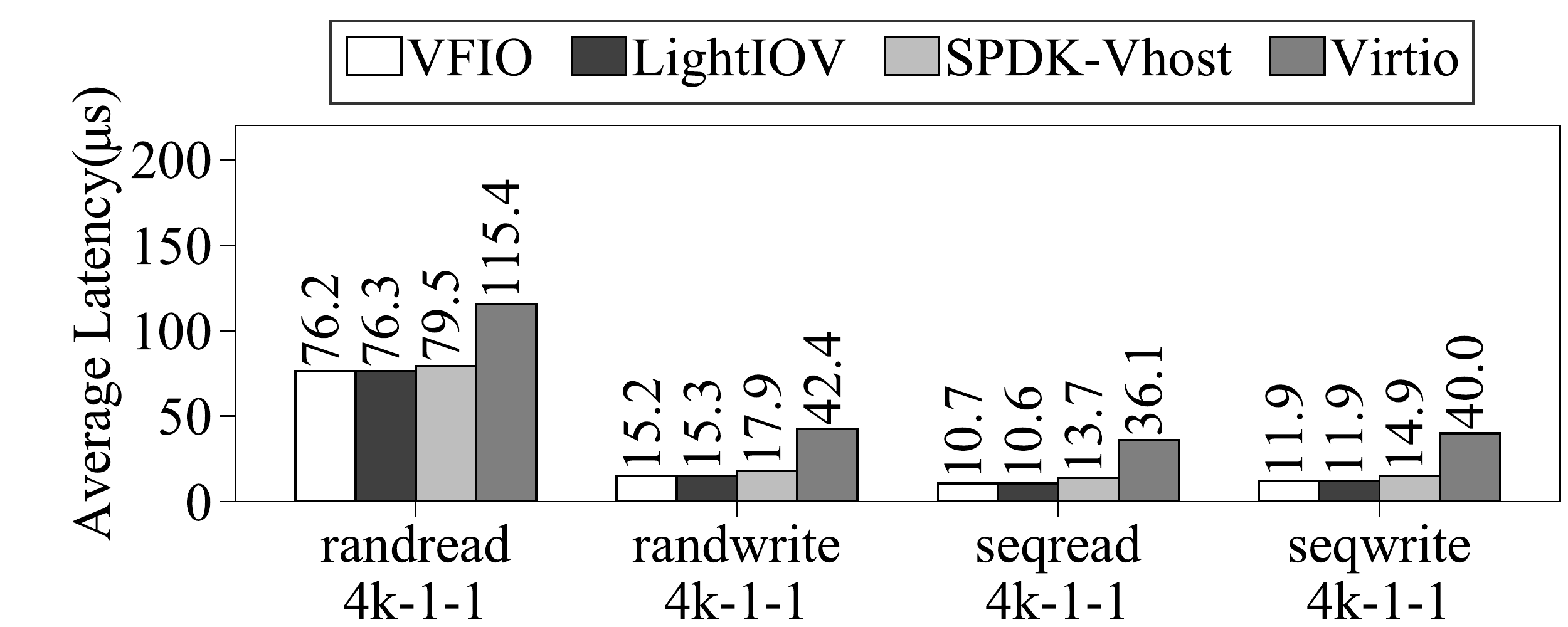}
  \subcaption{Average Latency}
  \label{fig:sig_lat}
\end{minipage}
\qquad
\begin{minipage}{\linewidth}
  \centering
   \setkeys{Gin}{width=\linewidth}   
   \includegraphics{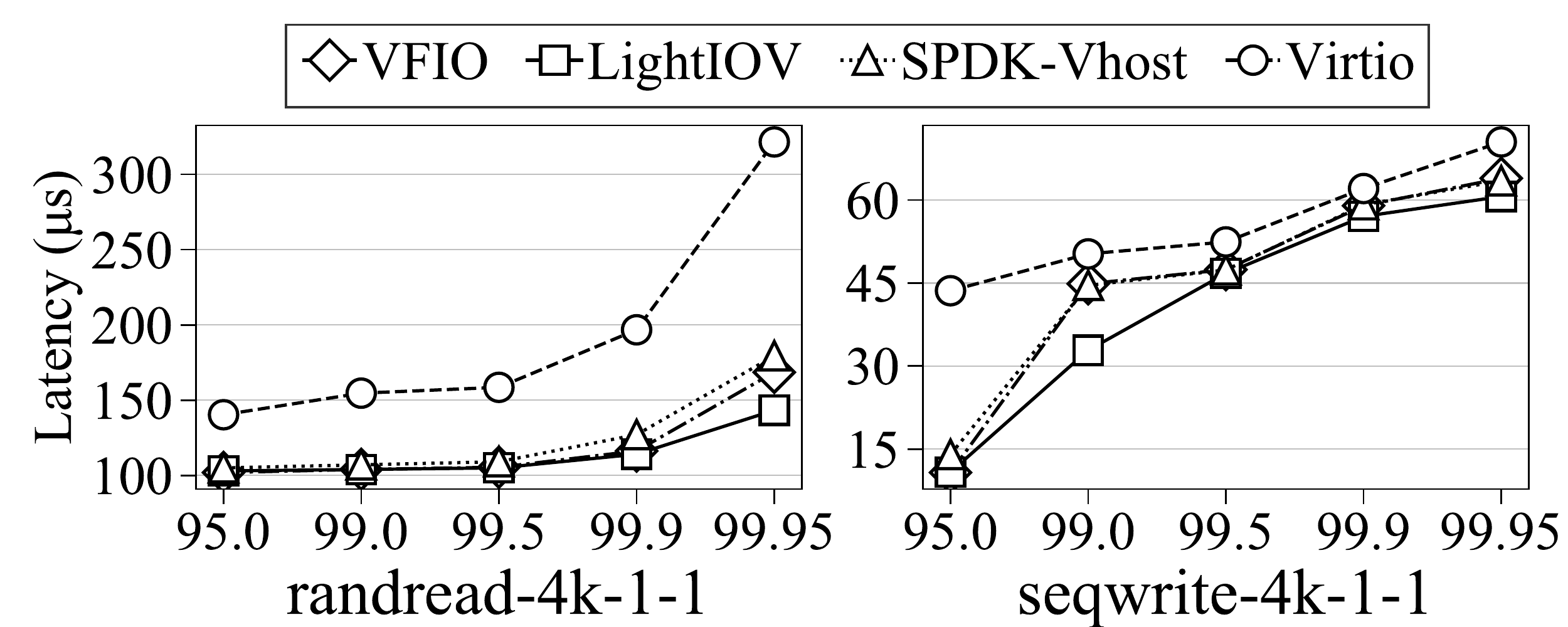}
  \subcaption{Tail Latency}
  \label{fig:sig_tail_lat}
\end{minipage}
\caption{Single-VM Latency with one SSD}
\vspace{-0.3cm}
\end{figure}

\section{Evaluation}
\label{evaluation}

We evaluate LightIOV and compare its I/O performance and scalability with other NVMe virtualization mechanism solutions, including VFIO, SPDK-Vhost, and virtio. We also evaluate the fairness of LightIOV on multiple VMs and run various synthetic benchmarks on real-world applications.  

\subsection{Experimental Setup}
{\bf System settings.} We deployed LightIOV on a host machine with two 48-core Intel Xeon Platinum 8163 CPUs. To avoid unpredictable fluctuations in test results from accessing devices across Non-Uniform Memory Access (NUMA) \cite{numa}, we used only one of the NUMA nodes. We used Intel SSD P4510 2T \cite{intel-p4510} as the NVMe device. 
For implementation, the LightIOV backend driver and LightIOV device are installed in the host, and the LightIOV frontend driver is installed in the guest. Both host and guest operating systems run on 64-bit CentOS with Linux kernel (version 4.19). In addition, we used a modified firecracker (version 0.14.0) as the hypervisor to manage VMs.

{\bf Compared virtualization solutions.}
We compared LightIOV with representative hardware-assisted and software-based solutions.
The performance of VFIO is close to the native disk, and its performance is the best among hardware-assisted solutions, so we use it as a baseline. 
For software-based solutions, we compare LightIOV with SPDK-Vhost and virtio. SPDK-Vhost refers to SPDK vhost-NVMe, which has the best performance of all SPDK vhost-target solutions.

{\bf Benchmarks and real-word application.}
  We used the fio \cite{fio} as our synthetic evaluation benchmark tool, which allows us to stress-test the disk by specifying specific I/O patterns (including random read and write, sequential read and write, etc.) and different I/O pressure (different threads and queue depth). Specifically, we took libaio as the fio engine and ran various test cases shown in Table~\ref{table:test cases}. We bypassed the guest OS kernel page cache  by setting the direct parameter to 1. In addition, RocksDB \cite{rocksdb}, a persistent key-value store for flash and RAM storage, was used to measure the performance of various virtualization solutions in real-world applications.

\begin{figure}[tbp]
\centering

\begin{minipage}{\linewidth}
  \centering
   \setkeys{Gin}{width=\linewidth}   
   \includegraphics{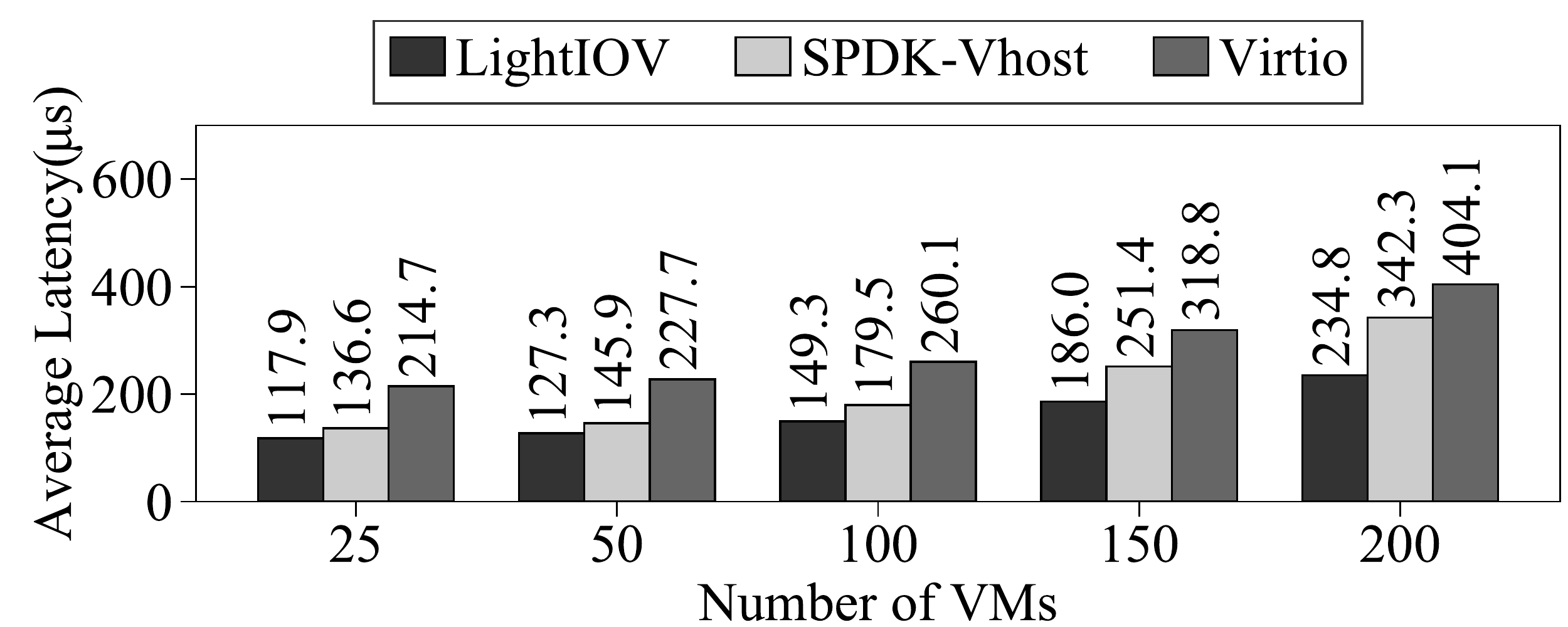}
  \subcaption{Average Latency of All VMs}
  \label{fig:mul_lat}
\end{minipage}

\begin{minipage}{\linewidth}
\centering
   \setkeys{Gin}{width=\linewidth}   
    \includegraphics{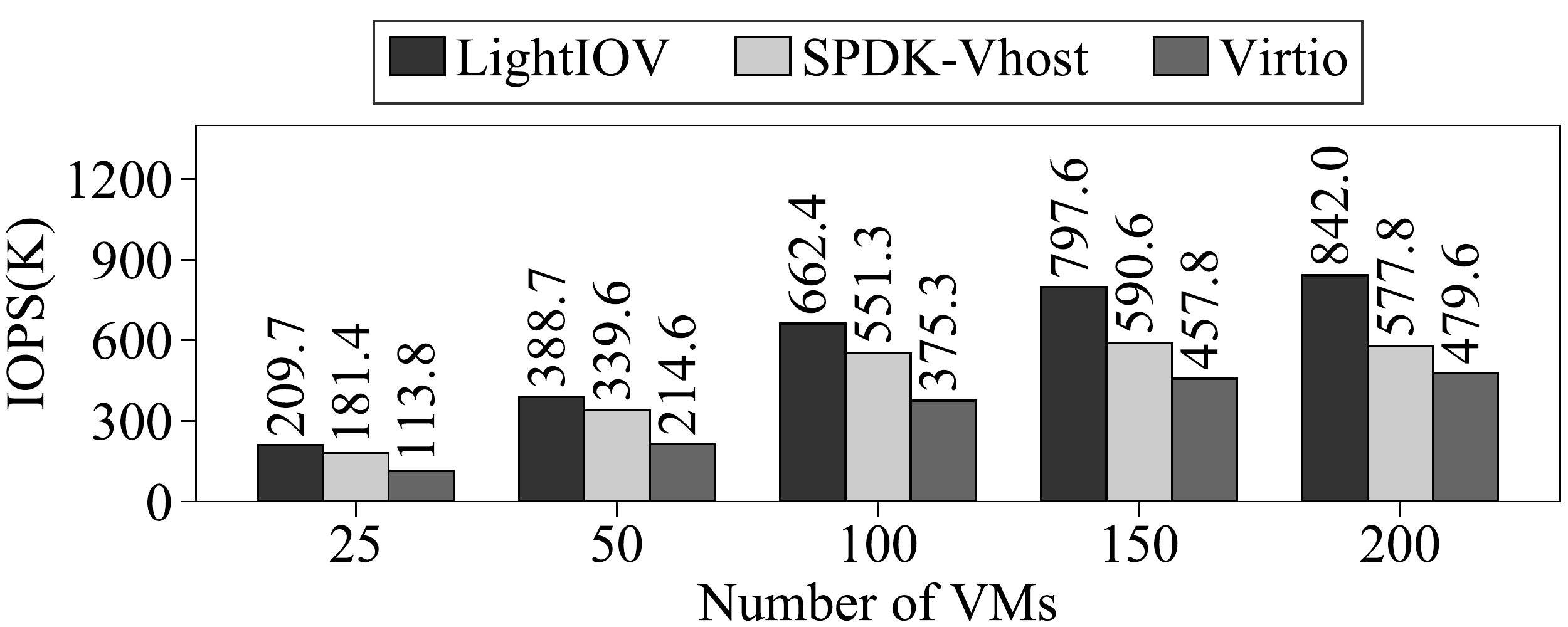}
  \subcaption{IOPS of All VMs}
  \label{fig:mul_iops}
\end{minipage}
\caption{I/O performance of multiple VMs running 4k-randread-1-1 on two SSDs}
\vspace{-0.4cm}
\end{figure}

\subsection{I/O performance}\label{single_vm} 
We evaluated the I/O performance of VFIO, LightIOV, SPDK-Vhost, and virtio with one SSD. We ran various test cases shown in Table~\ref{table:test cases} on a single VM, which is allocated with 4 CPU cores and 4GB of system memory. For SPDK-Vhost, we allocated one more CPU core for the SPDK vhost-target.

\textbf{Throughput.\label{subsec:throughput}}
Figure~\ref{fig:sig} demonstrate the throughput performance of a single VM. LightIOV can provide near-native performance in terms of IOPS and bandwidth. 

LightIOV can achieve 100.2\% and 97.6\% IOPS of VFIO in randread-4k-4-128 and randwrite-4k-4-128, respectively. VFIO passthrough the entire physical device to a VM, which can achieve near-native performance but sacrifice shareability. Whereas, LightIOV not only ensures near-native performance but also realizes device sharing among multiple VMs. 
Compared to SPDK, LightIOV has 6.2\% higher in IOPS randwrite-4k-4-128. SPDK-Vhost achieves performance close to LightIOV at the cost of consuming valuable CPU cores for polling.
Virtio suffers from severe performance degradation and shows the worst virtualized I/O performance. In randread-4k-4-128, LightIOV can achieve 335.8\% IOPS of virtio.

In the 128K sequential read/write test cases, all solutions reach the bandwidth limit of the storage device. LightIOV averagely achieves 99.4\% of VFIO, 100.9\% of SPDK, and 101\% of virtio. When processing large block requests, there is no significant performance difference among them.


\textbf{Latency.}
Figure~\ref{fig:sig_lat} depicts the average latency of single-thread and single-qd I/O workloads, and the tail latency of randread-4k-1-1 and seqwrite-4k-1-1 workloads are shown in Figure ~\ref{fig:sig_tail_lat}.

As seen from Figure~\ref{fig:sig_lat}, The average latency difference between LightIOV and VFIO is only less than 1\%  and can be negligible. It indicates that LightIOV can achieve the same low latency as VFIO. The latency of SPDK is up to 29.4\% higher than LightIOV in seqread-4k-1-1, because SPDK-Vhost has a longer I/O path than LightIOV. In LightIOV, VMs sends the I/O requests directly to the device without additional software processing. As a result, LightIOV achieves lower I/O latency than SPDK.
Virtio exhibits latency drawbacks due to its severe virtualization overhead, suffering from 51.2\%-236.1\% higher latency than LightIOV. In addition, we can see from Figure~\ref{fig:sig_tail_lat} that LightIOV exhibits the best tail latency, followed by VFIO and SPDK-Vhost, and finally, virtio.

To summarize, LightIOV can provide \textit{high performance} in terms of both throughput and latency. Its performance is comparable to VFIO, slightly better than SPDK-Vhost, and far better than virtio.



\begin{figure}[tbp]
\centering
\begin{minipage}{\linewidth}
\centering
   \setkeys{Gin}{width=\linewidth}   
    \includegraphics{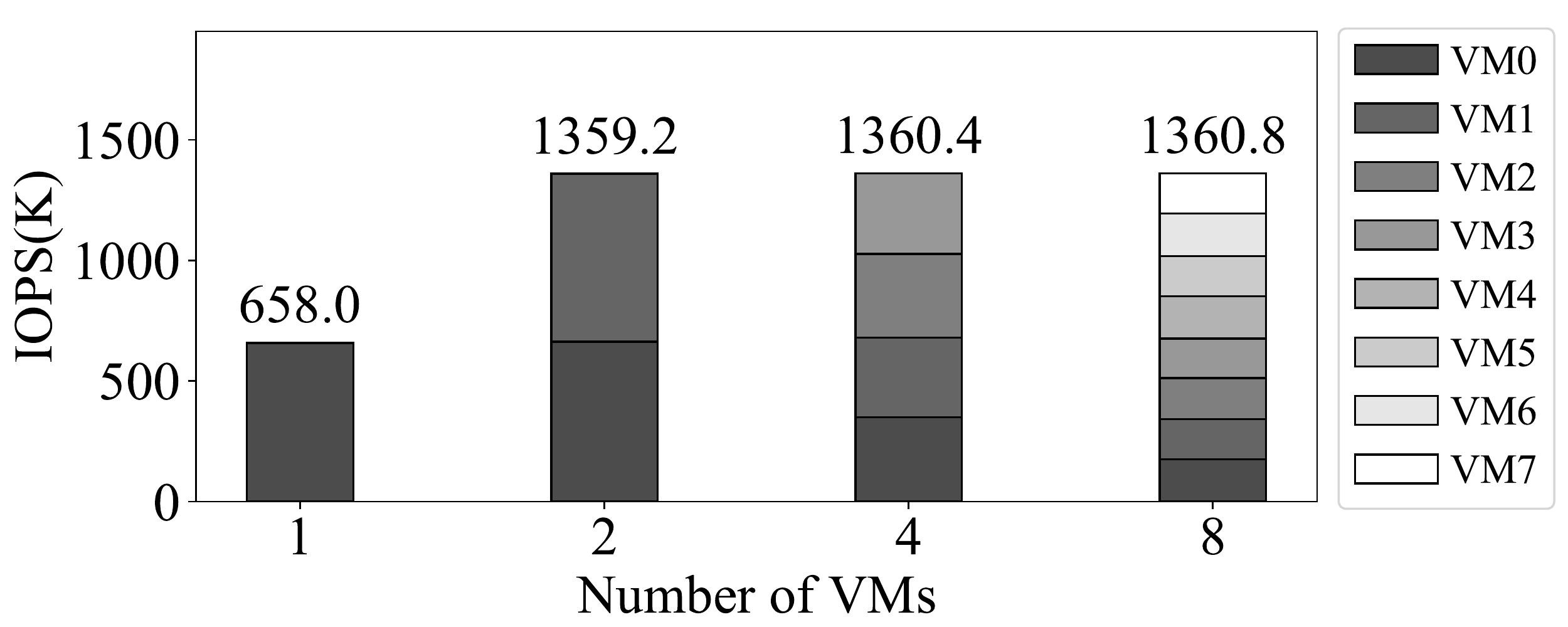}
  \subcaption{Overall IOPS of multiple VMs running on 2 SSDs}
  \label{fig:qos_iops}
\end{minipage}
\qquad
\begin{minipage}{\linewidth}
  \centering
   \setkeys{Gin}{width=\linewidth}   
   \includegraphics{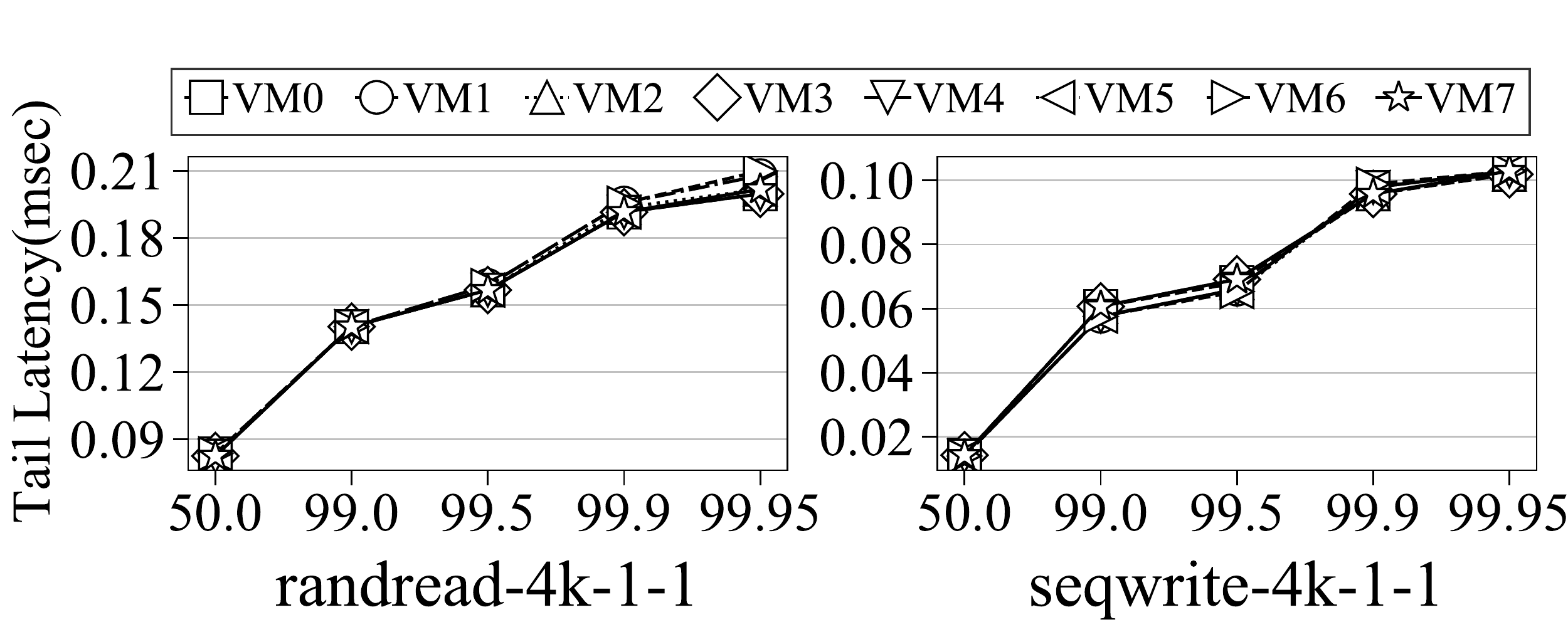}
  \subcaption{Tail latency distribution of eight VMs}
  \label{fig:qos_lat}
\end{minipage}
\caption{Fairness of LightIOV}
\vspace{-0.4cm}
\end{figure}

\subsection{Scalability}\label{scalability}
For the scalability test, we ran from 25 to 200 VMs on 2 SSDs and evaluated LightIOV, SPDK-Vhost, and virtio. 
VFIO directly assigns the device to a VM, which can not achieve device sharing by multiple VMs. Therefore it is not included in this subsection.
Four CPU cores are allocated for SPDK-Vhost polling because 4 cores are sufficient for 200 VMs on two SSDs, as shown in Figure \ref{fig:motivation}. We allocated one core for every 5 VMs. We configured the server with two Intel P4510 SSDs, and each VM owns 10G capacity. We ran 4k-randread-1-1 on these VMs and measured the average latency and overall throughput of all VMs. 



Figure \ref{fig:mul_lat} presents the latency performance of LightIOV, SPDK-Vhost, and virtio with different VM densities. Notably, LightIOV stands out with the lowest latency across all VM densities. As the VM density increases, the performance advantage of LightIOV over SPDK-Vhost becomes more significant, even though SPDK-Vhost consumes four more CPU cores than LightIOV. As the number of VMs escalates from 25 to 200, LightIOV provides 12.8\% to 31.4\% lower latency than SPDK-Vhost, which can be attributed to the I/O queues passthrough feature of LightIOV. In addition to outperforming SPDK-Vhost, LightIOV also displays a substantial latency advantage over virtio. Specifically, when running 25 VMs, LightIOV's latency is 45.1\% lower than that of virtio.

The results presented in Figure \ref{fig:mul_iops} demonstrate that LightIOV outperforms both SPDK-Vhost and virtio in terms of overall IOPS performance in high-density VMs. Specifically, when running 200 VMs, LightIOV provides 45.7\% and 75.6\% higher IOPS than SPDK-Vhost and virtio, respectively.

To summarize, LightIOV has superior I/O performance compared to software-baed NVMe virtualization solutions as the number of VMs increases, exhibiting \textit{high scalability}.

\begin{figure}[tbp]
\centering
\begin{minipage}{\linewidth}
\centering
   \setkeys{Gin}{width=\linewidth}   
    \includegraphics{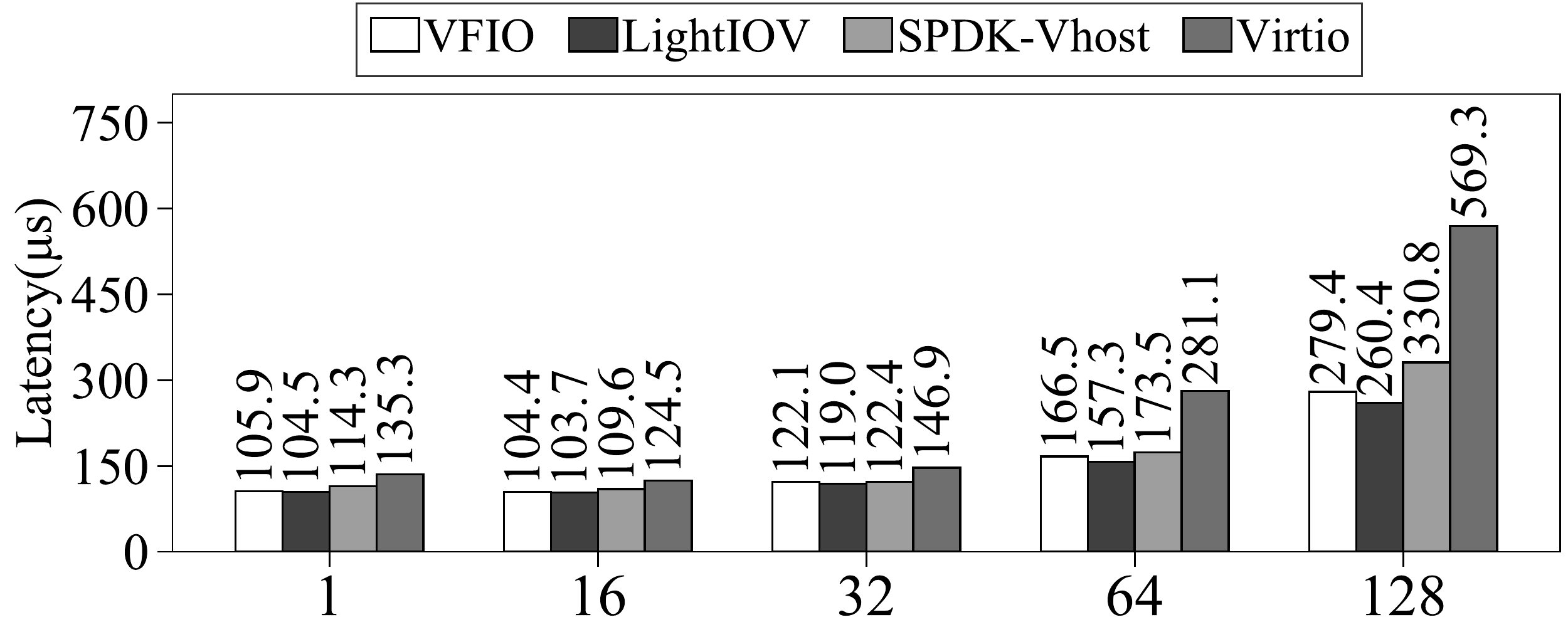}
  \subcaption{Latency of a single VM with increasing number of threads}
  \label{fig:rocksdb_single_threads_lat}
\end{minipage}
\qquad
\begin{minipage}{\linewidth}
  \centering
   \setkeys{Gin}{width=\linewidth}   \includegraphics{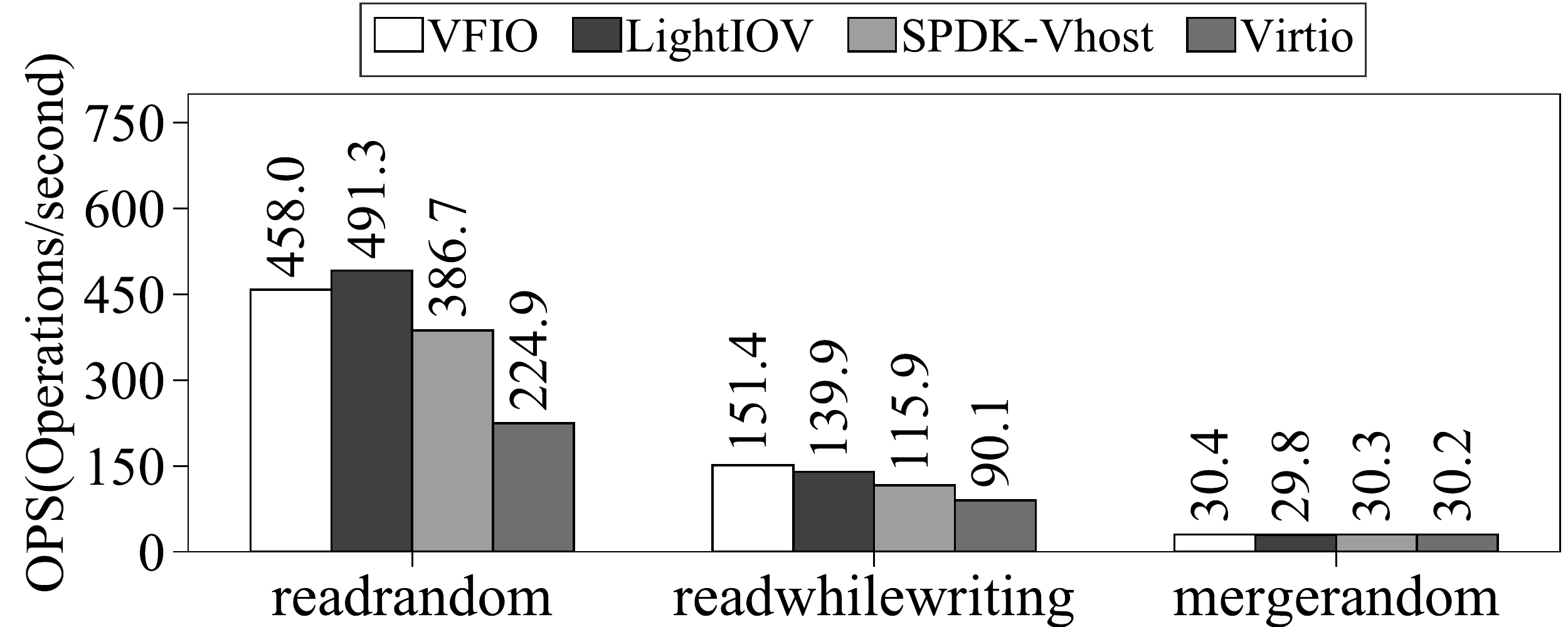}
  \subcaption{OPS of a single VM with different test cases using 128 threads}
  \label{fig:rocksdb_single_muljobs_128_iops}
\end{minipage}

\begin{minipage}{\linewidth}
  \centering
   \setkeys{Gin}{width=\linewidth}   
   \includegraphics{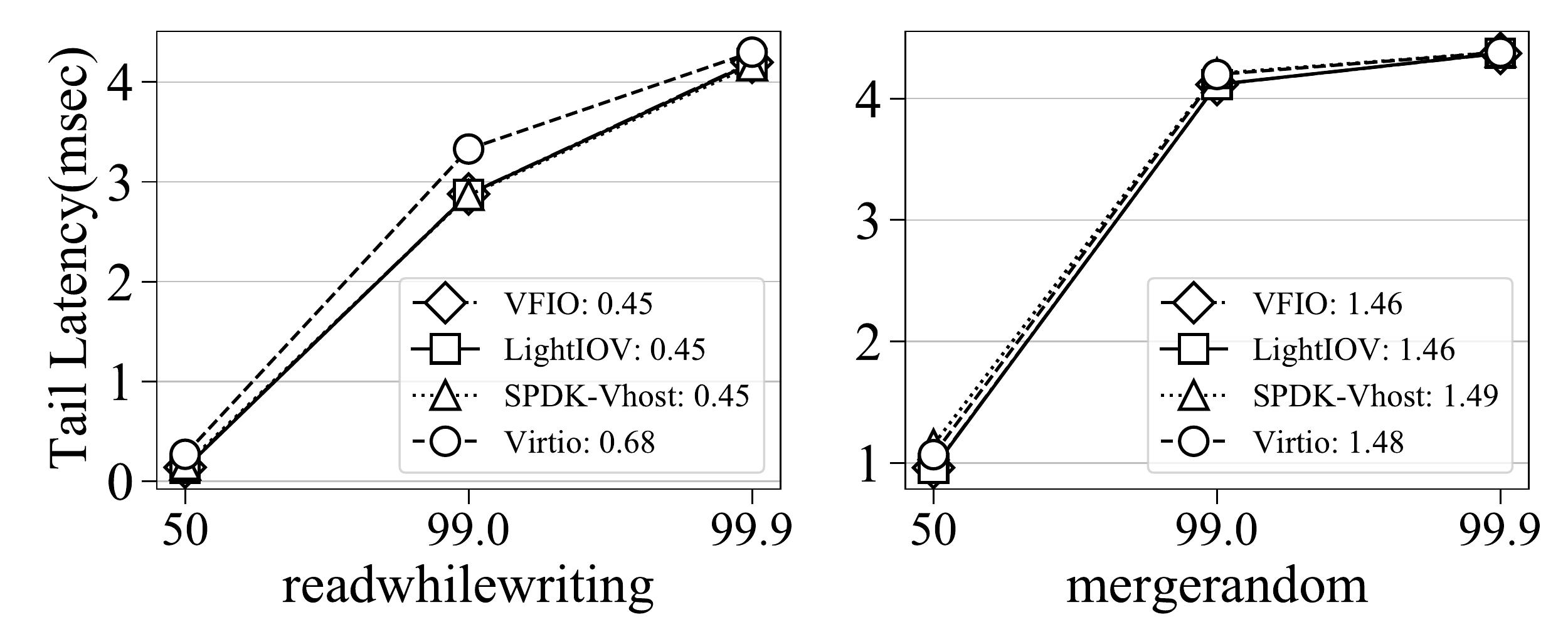}
  \subcaption{Tail latency of single VM with 1 thread}
  \label{fig:rocksdb_single_muljobs_1_tail_lat}
\end{minipage}

\caption{Single VM running RocksDB workloads}
\end{figure}


\subsection{Fairness}

This section evaluated fairness among multiple virtual machines when utilizing LightIOV. We assigned four CPU cores for each VM and ran 4k-randread-4-128 case.

Figure \ref{fig:qos_iops} shows the overall IOPS of LightIOV with 2 SSDs in multiple VMs. For this experiment, we utilized eight VMs, with VMs 0/2/4/6 utilizing one disk, while VMs 1/3/5/7 used the other disk.
When running one or two VMs, the overall IOPS reaches the limit of one SSD (658.0K IOPS) and two SSDs (1359.2K IOPS), respectively. However, when we increase the number of VMs to 4 and 8, the overall IOPS of all VMs remains at the IOPS limit of 2 SSDs (approximately 1360K IOPS). Additionally, the performance of each VM was evenly distributed, which demonstrated that the resources are efficiently utilized and distributed fairly among all VMs.

Moreover, we also examined the tail latency distribution of the eight VMs, as shown in Figure \ref{fig:qos_lat}. We observe that the tail latency of each VM is closely distributed, which indicated that the storage resources are not tilted toward some VMs. This further illustrates that all VMs were receiving an equal share of storage resources.

In summary, LightIOV can maintain the overall throughput with VM increasing and guarantee the \textit{fairness} of each VM.

\subsection{Real-world Application}

For the real-world application, we evaluated the performance of RocksDB, a persistent key-value store for flash and RAM storage, in both single and multiple VM scenarios with db\_bench initiating I/O requests. For the single-VM scenario, we assigned 4 CPU cores to each VM, while for the multi-VMs scenario, we assigned one core for every 5 VMs.
Specifically, we ran the "bulkload" case to insert 1 million data (with key size = 16 bytes, value size = 100 bytes, and total size = 1106.3 MB) into a database in sequential order. 

\textbf{Single-VM.} We conducted tests on various virtualization solutions while gradually increasing the number of threads in the readrandom test case from 1 to 128. Figure \ref{fig:rocksdb_single_threads_lat} illustrates that LightIOV outperforms VFIO and SPDK-Vhost in terms of average latency, and significantly surpasses virtio. Specifically, at 128 concurrent threads, LightIOV's average latency is 6.8\% lower than VFIO's and 21.3\% lower than SPDK-Vhost's, and its OPS is 54.3\% better than virtio.

We also conducted various workloads with 128 threads, including readrandom, readwhilewriting, and mergerandom on a single VM. The OPS (operations per second) are shown in Figure \ref{fig:rocksdb_single_muljobs_128_iops}, where higher values indicate better performance. LightIOV achieves performance close to VFIO in all test cases. Concretely, in the readrandom case, the OPS of LightIOV is 107.3\% of VFIO. Furthermore, LightIOV performs better than SPDK-Vhost and virtio in the readrandom and readwhilewriting cases. Specifically, the OPS of LightIOV in readrandom and readwhilewriting is 27.1\% and 20.7\% higher than SPDK-Vhost and 118.5\% and 55.2\% higher than virtio, respectively.

Figure \ref{fig:rocksdb_single_muljobs_1_tail_lat} presents the tail latency results for the readwhilewriting and mergerandom workloads performed on a single VM with a concurrency of 1. These results indicate that LightIOV offers similar latency performance to VFIO and SPDK-Vhost and outperforms virtio. LightIOV can achieve similar latency performance as VFIO with the advantage of queue passthrough. Virtio has the worst performance among the tested virtualization solutions due to its long I/O paths and context-switching overhead. 

Overall, the results demonstrate that LightIOV is a promising NVMe virtualization solution that offers a high throughput and low latency.


\begin{figure}[tbp]
\centering
\begin{minipage}{\linewidth}
\centering
   \setkeys{Gin}{width=\linewidth}   
    \includegraphics{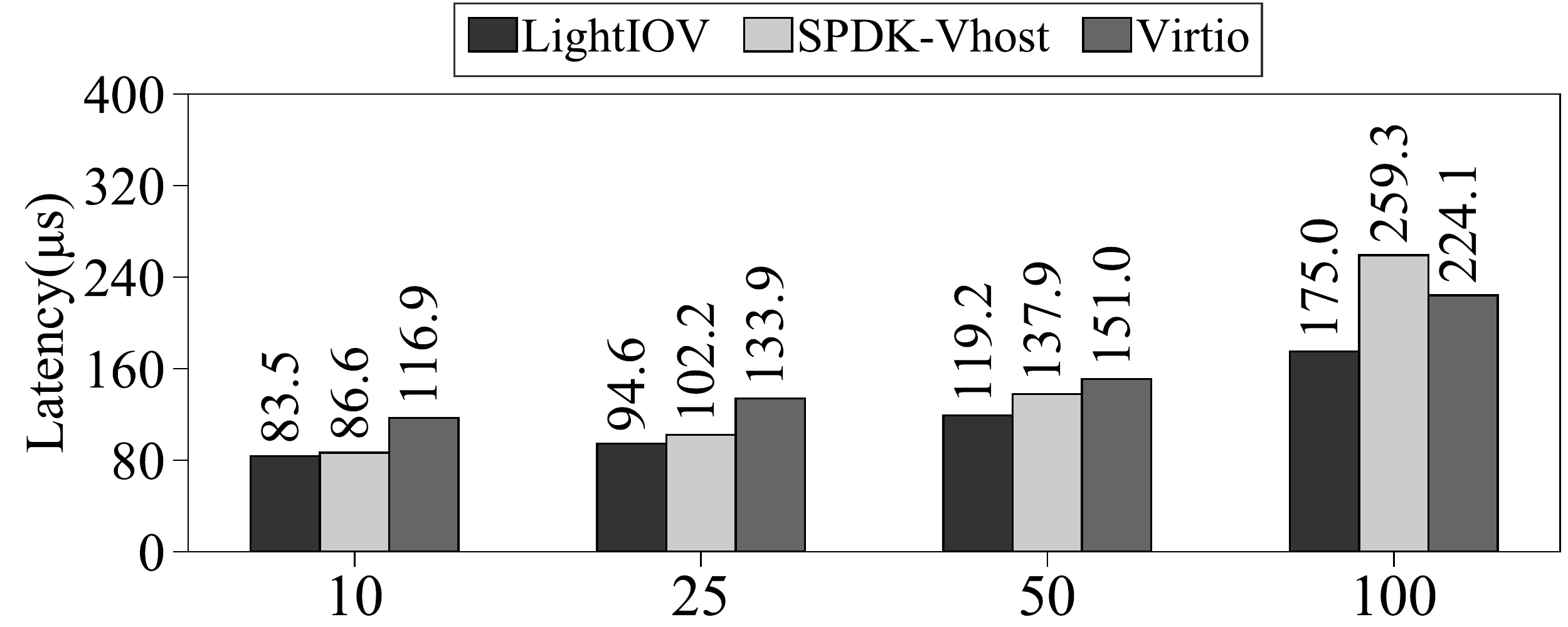}
  \subcaption{Average latency of increasing VMs running readrandom with 1 thread}
  \label{fig:rocksdb_mul_vms_1_lat}
\end{minipage}
\qquad
\begin{minipage}{\linewidth}
  \centering
   \setkeys{Gin}{width=\linewidth}   
   \includegraphics{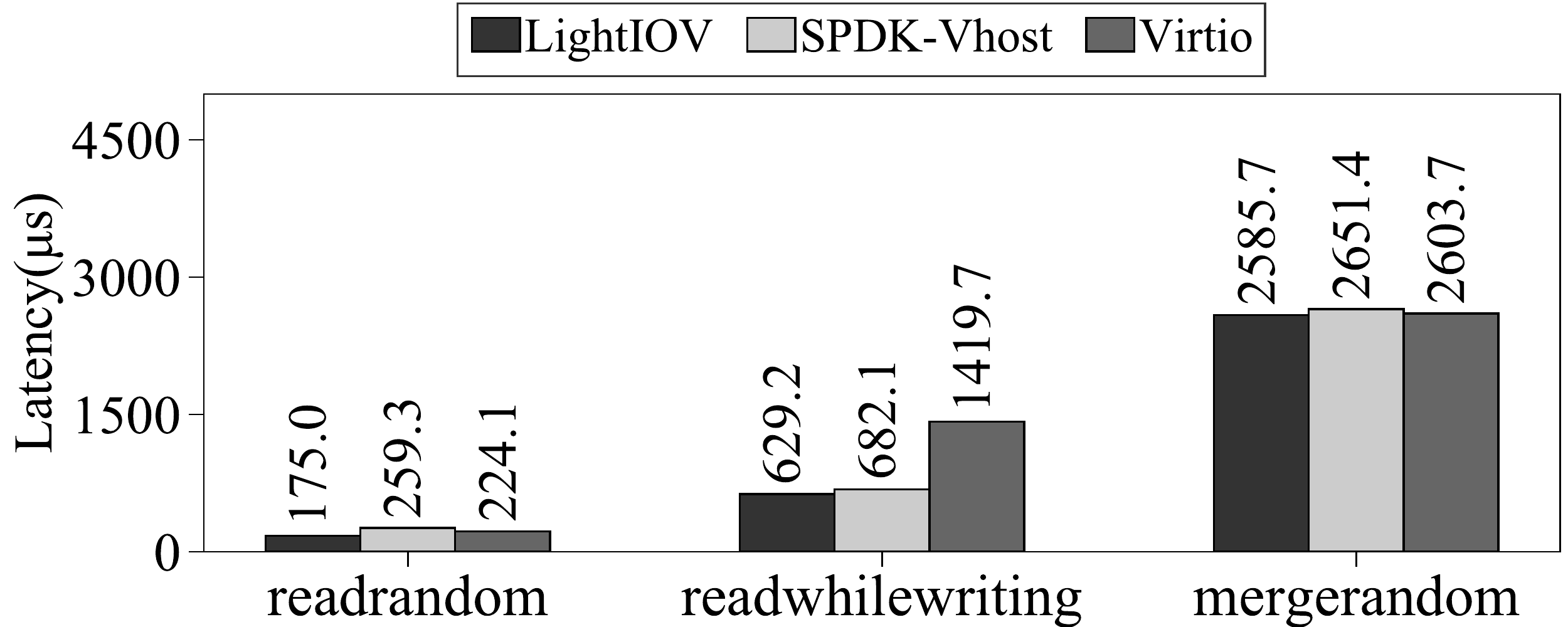}
  \subcaption{Average latency of 100 VMs running various workloads with 1 thread}
  \label{fig:rocksdb_mul_muljobs_1_lat}
\end{minipage}
\caption{Multiple VMs running RocksDB workloads}
\vspace{-0.4cm}
\end{figure}


\textbf{Multi-VMs.} To begin with, we conducted a readrandom workload with a concurrency of 1 on multiple VMs, gradually increasing the number of VMs up to 100. As Figure \ref{fig:rocksdb_mul_vms_1_lat} demonstrates, LightIOV consistently maintains its performance advantage as the number of VMs increases. As the number of VMs rises from 10 to 100, LightIOV's total latency is consistently 3.6\% to 32.5\% lower than that of SPDK. Notably, when the VM count reaches 25, LightIOV outperforms virtio by 29.3\%, showcasing its superior scalability. These results demonstrate that LightIOV is a highly scalable solution that maintains its performance advantage in real-world workloads.

Then, we conducted various workloads on 100 VMs, and each VM only runs one thread. The average latency of each VM is shown in Figure \ref{fig:rocksdb_mul_muljobs_1_lat}.  Specifically, in the readrandom case, the latency of LightIOV is 32.5\% lower than SPDK-Vhost. 
When compared with virtio, LightIOV also demonstrated better performance, with 21.9\% and 55.7\% lower latency in readrandom and readwhilewriting cases, respectively.

The results obtained from single and multiple VMs demonstrate that LightIOV outperforms both SPDK-Vhost and virtio in terms of OPS and latency when it comes to real-world applications.

\section{Related Work}
\noindent\textbf{Software-based NVMe virtualization.}
Virtio \cite{virtio} is a series of well-maintained Linux drivers for general I/O device virtualization, but it suffers from significant performance degradation. 
To address this issue, advanced polling-based approaches have been proposed to assist the hypervisor with virtual device I/O handling. SPDK vhost-NVMe \cite{spdk_2018} is an I/O service target that relies on user space NVMe drivers to eliminate unnecessary VM\_Exit overhead and shrink the I/O execution stack in the host OS. MDev-NVMe \cite{mdevnvme} is a mediated passthrough mechanism that utilizes a polling mode, and further research \cite{mdev-poll} has proposed an adaptive polling mode for MDev-NVMe to achieve better performance or save CPU resources for better scalability. FinNVMe \cite{mdev-throughput} is a high-throughput storage virtualization management solution for an NVMe storage device that works in a workload-aware manner among multi-tenant VMs. 
Finally, Direct-Virtio \cite{kim_direct-virtio_2021} makes use of user-level storage direct access to avoid duplicated I/O stack overhead with polling methods. However, these polling-based solutions come at the cost of high CPU overhead.

\textbf{Hardware-assisted NVMe virtualization.}
 VFIO \cite{vfiouser} provides exclusive device assignment to a single VM, which goes against the shareability concept of virtualization.
SR-IOV \cite{sriov} enables multiple VMs to share Physical Functions (PFs) and Virtual Functions (VFs) and achieve high performance.
FVM \cite{kwon_fvm_nodate} is a solution that builds on SR-IOV by offloading the virtualization layer to an FPGA card, which can directly manage the physical devices. This approach can reduce the overhead associated with software-based virtualization and enable high performance for demanding workloads.
LeapIO \cite{leapio} offloads the entire storage stack to an ARM SoC with support for both local and remote storage.
Scalable IOV \cite{siov} is a novel approach to hardware-assisted I/O virtualization that aims to provide highly scalable and high-performance virtualization for I/O devices, which rely on the advanced hardware features and optimizations to enable efficient virtualization of I/O devices at scale.
BM-Store \cite{bmstore} relies on an FPGA-based BMS-Engine and an ARM-based BMS-Controller to achieve transparent and high-performance virtual local storage for bare-metal clouds.

\section{Future Work}
LightIOV allows VMs to access NVMe devices directly, providing high performance, low overhead, flexibility, and scalability with NVMe I/O queue passthrough. However, due to the internal request processing mechanism of the SSD controller, performance interference between virtual machines is inevitable, despite each VM having exclusive I/O queues and unique LBAs. After retrieving the I/O requests from the host memory, the SSD controller goes through further processing through the host-interface logic (HIL), the flash translation layer (FTL), and flash channel controllers (FCCs) before sending the I/O requests down to the back-end address space. Therefore, requests from different VMs can interfere with each other inside the SSD controller, leading to reduced performance. For example, when two VMs share an SSD, one running a light workload (with a small block size and low queue depth) while the other runs a heavy workload (with a large block size and high queue depth), the former's performance will be severely impacted. In the future, we plan to enhance the NVMe controller at the firmware level by binding I/O queues with backend queues, dedicated channels or dies, and optimizing cache management to reduce performance interference between different VMs.

Furthermore, NVMe devices often require quality of service (QoS) capabilities to differentiate I/O read and write services based on upper-layer business requirements, such as a 99.99\% tail latency. To address this challenging issue, we aim to associate shareable hardware resources, such as NVMe Namespace or LBA ranges, I/O Queues, and PRP/SGLs, within the NVMe device controller to form prioritized NVMe resource domains for virtual machines. This approach will enable different virtual machines to achieve varying levels of I/O priority services, ensuring that each virtual machine can receive the necessary resources to meet its unique business needs.
\section{Conclusion}
In this paper, we propose LightIOV, a novel high-performance and scalable software-based NVMe virtualization mechanism that can easily scale up to tens of thousands of VMs. The key idea of LightIOV is NVMe I/O queues passthrough, which enables VMs directly access the NVMe I/O queues. LightIOV provides full virtualized NVMe devices by combining virtualized control resources (including PCIe configuration, BAR space, and admin queue) with data resources (including NVMe I/O queues, doorbell registers,  interrupt resources, and LBAs). Furthermore, with IOMMU support, LightIOV achieves host-bypassing DMA transactions and interrupts processing to enhance NVMe virtualization performance. In this way, LightIOV achieves high performance and high scalability while not consuming valuable CPU resources and not requiring special hardware support.


\bibliographystyle{ACM-Reference-Format}
\bibliography{references}










\end{document}